\documentclass[aps,                
               prc,                
               showpacs,           
               showkeys,           
               twocolumn,          
               10pt,               
               twoside,            
               floatfix,           
               superscriptaddress] 
               {revtex4-1}

\usepackage{mathptmx}         
\usepackage[T1]{fontenc}
\usepackage{geometry}         
\usepackage{fancyhdr}         
\usepackage{graphicx}         
\usepackage[svgnames]{xcolor} 
\usepackage{hyperref}         
\usepackage{amsmath}          
\usepackage{amssymb}          
\usepackage{amsfonts}         
\usepackage{amstext}          
\usepackage{mathrsfs}         
\usepackage{array}            
\usepackage{textcomp}         
\usepackage{cancel}           

\hypersetup{colorlinks=true,linkcolor=Blue,citecolor=Blue,urlcolor=Blue}

\DeclareMathAlphabet{\mathcal}{OMS}{cmsy}{m}{n}

\DeclareGraphicsExtensions{.pdf}

\fancypagestyle{firststyle}
{%
  \fancyhf{}
  \fancyhead[C]{PHYSICAL REVIEW C {\bf 88}, 045204 (2013)} 
  \fancyfoot[L]{\small 0556-2813/2013/88(4)/045204(13)} 
  \fancyfoot[C]{\small 045204-\thepage} 
  \fancyfoot[R]{\small \textcopyright 2013 American Physical Society} 
}

\newcolumntype{I}{!{\vrule width 1.1pt}}

\newlength{\savedwidth}

\newcommand{\whline}%
{%
  \noalign{\global\setlength{\savedwidth}{\arrayrulewidth}}%
  \noalign{\global\setlength{\arrayrulewidth}{1.1pt}}\hline%
  \noalign{\global\setlength{\arrayrulewidth}{\savedwidth}}%
}

\newcommand{\wcline}[1]%
{%
  \noalign{\global\setlength{\savedwidth}{\arrayrulewidth}}%
  \noalign{\global\setlength{\arrayrulewidth}{1.1pt}}\cline{#1}%
  \noalign{\global\setlength{\arrayrulewidth}{\savedwidth}}%
}

\begin{document}

  \title{\texorpdfstring{Transport coefficients from the Nambu-Jona-Lasinio model for $SU(3)_f$}{Transport coefficients from the Nambu-Jona-Lasinio model for SU(3)f}}

\author{R.~Marty}
\email[Email : ]{marty@fias.uni-frankfurt.de}
\affiliation{\begin{minipage}[c]{0.95\textwidth}Frankfurt Institut for Advanced Studies, Johann Wolfgang Goethe Universit\"at, Ruth-Moufang-Str. 1, 60438 Frankfurt am Main, Germany\end{minipage}}
\affiliation{\begin{minipage}[c]{0.91\textwidth}Institut for Theoretical Physics, Johann Wolfgang Goethe Universit\"at, Max-von-Laue-Str. 1, 60438 Frankfurt am Main, Germany\end{minipage}}

\author{E.~Bratkovskaya}
\affiliation{\begin{minipage}[c]{0.95\textwidth}Frankfurt Institut for Advanced Studies, Johann Wolfgang Goethe Universit\"at, Ruth-Moufang-Str. 1, 60438 Frankfurt am Main, Germany\end{minipage}}
\affiliation{\begin{minipage}[c]{0.91\textwidth}Institut for Theoretical Physics, Johann Wolfgang Goethe Universit\"at, Max-von-Laue-Str. 1, 60438 Frankfurt am Main, Germany\end{minipage}}

\author{W.~Cassing}
\affiliation{\begin{minipage}[c]{0.73\textwidth}Institut f\"ur Theoretische Physik, Universit\"at Gie\ss{}en, Heinrich-Buff-Ring 16, 35392 Gie\ss{}en, Germany\end{minipage}}

\author{J.~Aichelin}
\affiliation{\begin{minipage}[c]{0.98\textwidth}Subatech, UMR 6457, IN2P3/CNRS, Universit\'e de Nantes, \'Ecole des Mines de Nantes, 4 rue Alfred Kastler, 44307 Nantes cedex 3, France\end{minipage}\\
\normalfont{(Received 22 May 2013; revised 12 June 2013; published 26 June 2013)}\vspace{2.5mm}}

\author{H.~Berrehrah}
\affiliation{\begin{minipage}[c]{0.95\textwidth}Frankfurt Institut for Advanced Studies, Johann Wolfgang Goethe Universit\"at, Ruth-Moufang-Str. 1, 60438 Frankfurt am Main, Germany\end{minipage}}

\pacs{11.10.Wx, 51.30.+i, 51.20.+d, 25.75.Nq%
\hfill DOI:\href{http://dx.doi.org/10.1103/PhysRevC.88.045204}{10.1103/PhysRevC.88.045204}}



\begin{abstract}
We calculate the shear $\eta(T)$ and bulk viscosities $\zeta(T)$ as well as the electric conductivity $\sigma_e(T)$ and heat conductivity $\kappa(T)$ within the Nambu-Jona-Lasinio model for three flavors as a function of temperature as well as the entropy density $s(T)$, pressure $P(T)$ and speed of sound $c_s^2(T)$. We compare the results with other models such as the Polyakov-Nambu-Jona-Lasinio (PNJL) model and the dynamical quasiparticle model (DQPM) and confront these results with lattice QCD data whenever available. We find the NJL model to have a limited predictive power for the thermodynamic variables and various transport coefficients above the critical temperature whereas the PNJL model and DQPM show acceptable results for the quantities of interest.
\end{abstract}

\maketitle

  \section{Introduction}
\thispagestyle{firststyle}
\vskip -3mm
For small momentum transfers, quantum chromo-dynamics (QCD), the fundamental theory of strong interactions, can presently only be solved on a finite Euclidean lattice in thermodynamic equilibrium. Even this solution requires computer power in teraflops and many of the most interesting questions such as the mass of hadrons at finite temperature and chemical potential as well as the properties of the constituents of deconfined matter and their interactions cannot be addressed presently accurately, either due to limited computer power or due to conceptual
problems, in particular at finite quark chemical potential $\mu$.

Therefore one is strongly motivated to look for simpler models that incorporate essential features of QCD but are mathematically tractable. A candidate for this is the Nambu-Jona-Lasinio (NJL) model which is constructed in a way that its Lagrangian shares the symmetries of QCD --which are also observed in nature-- but also the breaking of these symmetries. One of the most important of these symmetries is chiral symmetry, which is essential for the understanding of the hadron properties and their excitation spectra. We recall that the breaking of this symmetry is responsible for the dynamical creation of the fermion masses.

Whereas in former times the NJL approach has been mainly used to describe and to understand hadron properties at finite density and temperature, in recent times the interest has been extended to a systematic investigation of the properties and  the interactions of quarks in deconfined matter, especially at temperatures above the critical temperature $T_c$ for the phase transition to deconfined matter. Without introducing additional parameters elastic scattering and hadronization cross sections for quarks and antiquarks have been calculated \cite{Rehberg1996b} which have more recently been used to model the expansion of a quark-antiquark plasma with initial conditions obtained for nucleus-nucleus collisions at the Relativistic Heavy Ion Collider (RHIC) \cite{Marty2012}. In order to compare this approach with other transport models, which study the expansion of the plasma created in ultra relativistic heavy-ion collisions, it is useful to calculate and compare thermodynamic properties as well as transport coefficients in equilibrium as a function of the temperature $T$.

It is the purpose of this article to evaluate these quantities in the NJL model and to compare them with similar approaches such as the Polyakov-Nambu-Jona-Lasinio (PNJL) \cite{Ratti2007} and the dynamical-quasiparticle model (DQPM) \cite{Peshier2004,Peshier2005,Cassing2009a}. The latter one is presently used in an independent transport approach which describes the time evolution of heavy-ion collisions from the creation of the partonic plasma until the registration of the final hadrons in the detectors, i.e., the parton-hadron string dynamics (PHSD) approach \cite{Cassing2008,Bratkovskaya2011,Cassing2009b}. Both models are quite different as far as the temperature dependence of the parton masses and the hadronization dynamics are concerned. It is therefore of interest to study whether these different ingredients lead to a different dynamics of the system. We will compare, as much as possible, all results also with available lattice QCD data.

\begin{figure*}
  \begin{center}
    \includegraphics[width=8.5cm]{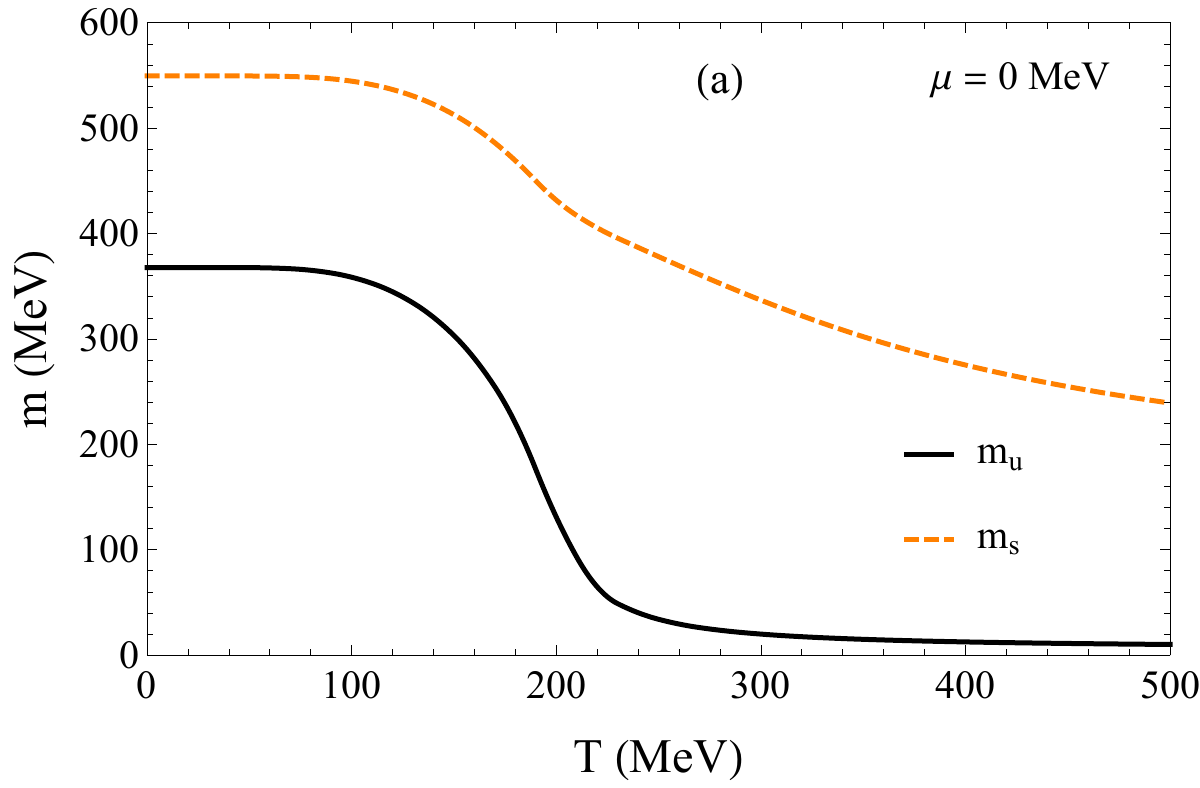} \ \ \
    \includegraphics[width=8.5cm]{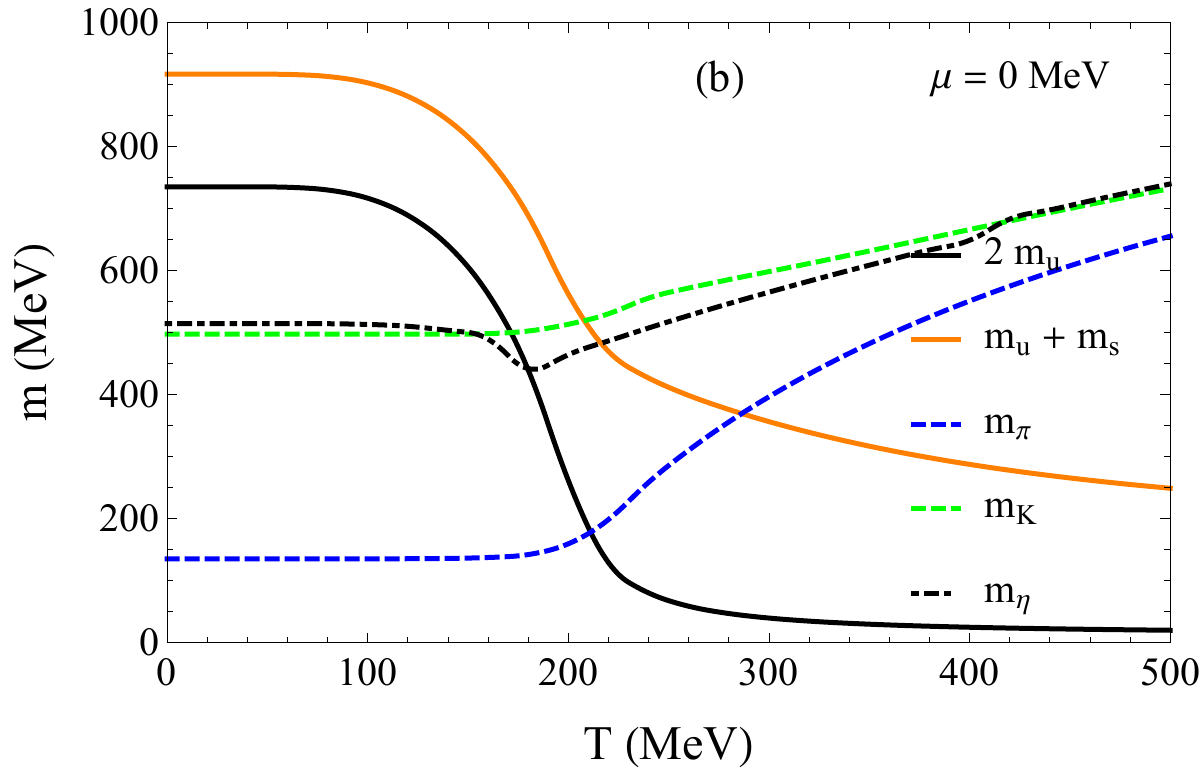}
  \end{center}
  \vskip -5mm
  \caption{(Color online) Masses of quarks (a) and pseudoscalar mesons (b) as a function of the temperature $T$ in the NJL model.\label{MassNJL}}
\end{figure*}

\section{The Nambu--Jona-Lasinio model}
\vskip -3mm
Before quantum chromo-dynamics (QCD) was formulated in 1973, Y. Nambu and G. Jona-Lasinio advanced a model for strongly interacting fermions \cite{Nambu1961} --nucleons in their case-- by a four-point interaction. The use of the NJL model was extended to quarks and antiquarks only at the beginning of the 1990s, assuming the gluon degrees of freedom to be integrated out. The NJL Lagrangian presently used for our purpose for $SU(3)_f$ is taken from Ref.
\cite{Klevansky1992}:
\begin{equation}
  \begin{aligned}
    \mathscr{L}_{NJL} =&\bar{\psi} \left( i \partial\!\!\!\!/- m_0 \right) \psi\\
             &+ \ G \sum_{a=0}^8 \left[   \left( \bar{\psi}            \lambda^a \psi \right)^2
                                        + \left( \bar{\psi} i \gamma_5 \lambda^a \psi \right)^2 \right] \\
             &- \ K \left[   \mathrm{det} \bar{\psi} \left( 1 - \gamma_5 \right) \psi
                           + \mathrm{det} \bar{\psi} \left( 1 + \gamma_5 \right) \psi \right],
  \end{aligned}
  \label{NJL_lagrangian}
\end{equation}
where $\psi = u,d,s$ is the quark field with three flavors ($N_f = 3$) and three colors ($N_c = 3$), $\lambda^a$ are the flavor $SU(3)_f$ Gell-Mann matrices ($a = 0, 1, \dots , 8$), and $G$ and $K$ are coupling constants. The first line is the  Lagrangian of freely moving quarks with a constituent mass $m_0$. The second(third) line of Eq. \eqref{NJL_lagrangian} describes the four-quark(six-quark) interaction. The symmetry group of the NJL model is the same as that of  QCD:
\begin{equation}
  \text{U}(3)_L \otimes \text{U}(3)_R =
  \text{SU}_V(3) \otimes \text{SU}_A(3) \otimes \text{U}_V(1) \otimes \text{U}_A(1).
\end{equation}
Chiral symmetry is spontaneously broken in vacuum and explicitly broken by the bare mass $m_0$ of quarks in Eq. \eqref{NJL_lagrangian}. The well known axial anomaly $U_A(1)$ is broken by the last term of Eq. \eqref{NJL_lagrangian} which relates to the 't Hooft mixing.

\begin{table} [!t]
  \caption{Parameters for the NJL model from Ref. \cite{Rehberg1996b}.\label{paramNJL}}
  \begin{ruledtabular}
    \begin{tabular}{ c c c c c }
      $\Lambda$ & $G\Lambda^2$ & $K\Lambda^5$ & $m_{0q}$ & $m_{0s}$ \\
      \hline
      \noalign{\vskip 3pt}
      602.3 MeV & 1.835 & 12.36 & 5.5 MeV & 140.7 MeV \\
    \end{tabular}
  \end{ruledtabular}
\end{table}

From this Lagrangian we can get effective masses for quarks. In the mean-field approximation they are given by
\begin{equation}
  m_i = m_{0i} - 4 G \langle\langle \bar{\psi_i}\psi_i \rangle\rangle
               + 2 K \langle\langle \bar{\psi_j}\psi_j \rangle\rangle
                     \langle\langle \bar{\psi_k}\psi_k \rangle\rangle,
  \label{massNJL}
\end{equation}
with $i \ne j \ne k$ denoting the three possible flavors of quarks, and $\langle\langle \bar{\psi}\psi \rangle\rangle$ being the chiral condensate in the mean field limit,
\begin{equation}
  \langle\langle \bar{\psi}_i\psi_i \rangle\rangle
  = - 2 N_c \ \int_0^\Lambda \frac{d^3 p}{(2\pi)^3}
  \frac{m_i}{E_{i \bf p}} [1 - f_q - f_{\bar q}],
  \label{NJL_condensate}
\end{equation}
with $E_{\bf p} = \sqrt{{\bf p}^2 + m^2}$, $\Lambda$ being the NJL cutoff, and the Fermi-Dirac distribution
\begin{equation}
  \begin{aligned}
    f_q       ({\bf p},T,\mu)&= \{\exp[(E_{\bf p}-\mu) / T] + 1\}^{-1},\\
    f_{\bar q}({\bf p},T,\mu)&= \{\exp[(E_{\bf p}+\mu) / T] + 1\}^{-1}.
  \end{aligned}
  \label{fermidis}
\end{equation}

\begin{figure*}
  \begin{center}
    \includegraphics[width=8.5cm]{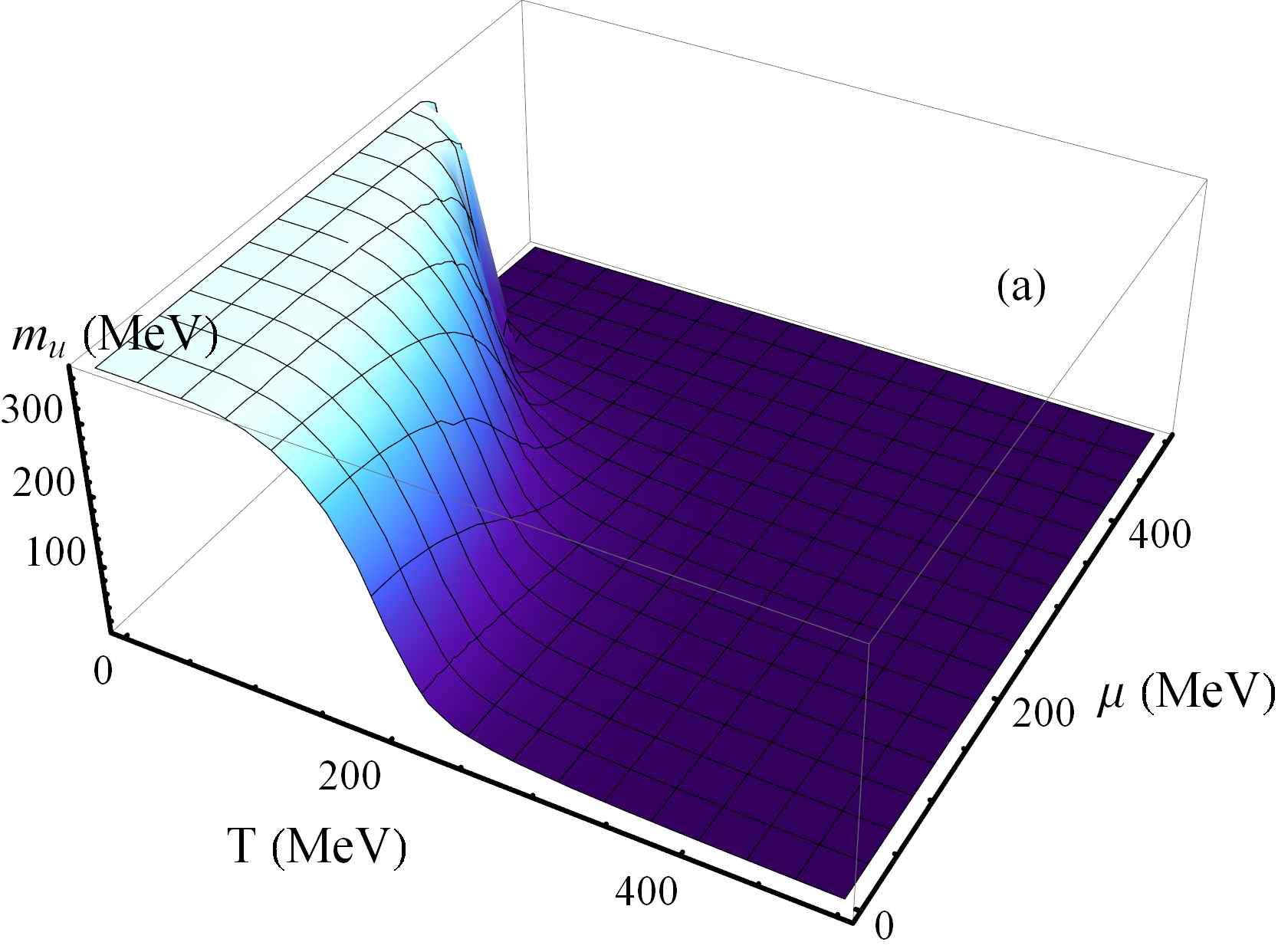} \ \ \
    \includegraphics[width=8.5cm]{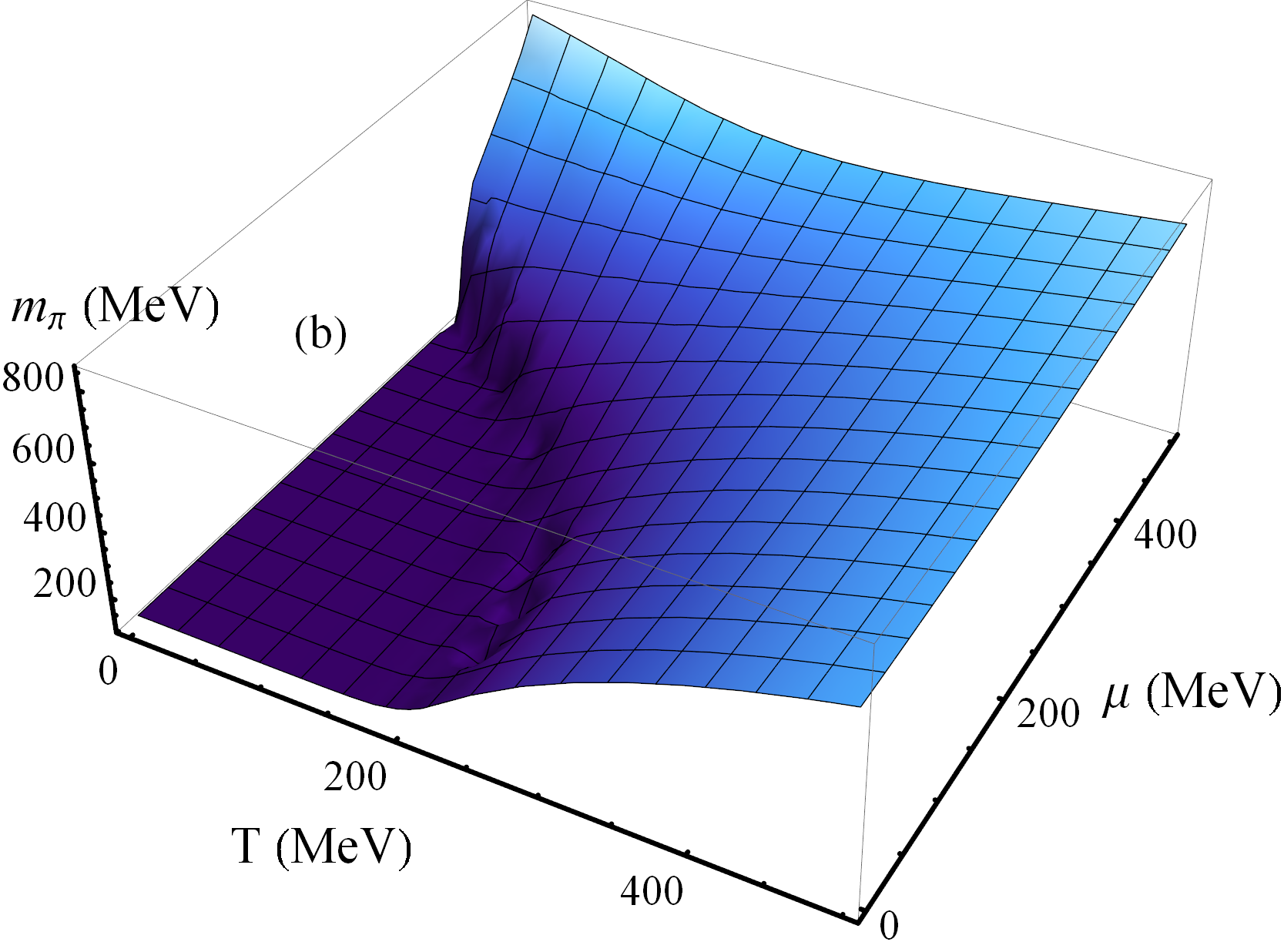}
  \end{center}
  \vskip -3mm
  \caption{(Color online) Masses of $u$ quarks (a) and $\pi$ mesons (b) as a function of the temperature $T$ and the quark chemical potential $\mu$ in the NJL model.\label{Mass3D}}
\end{figure*}

\begin{figure*}
  \begin{center}
    \includegraphics[width=8.5cm]{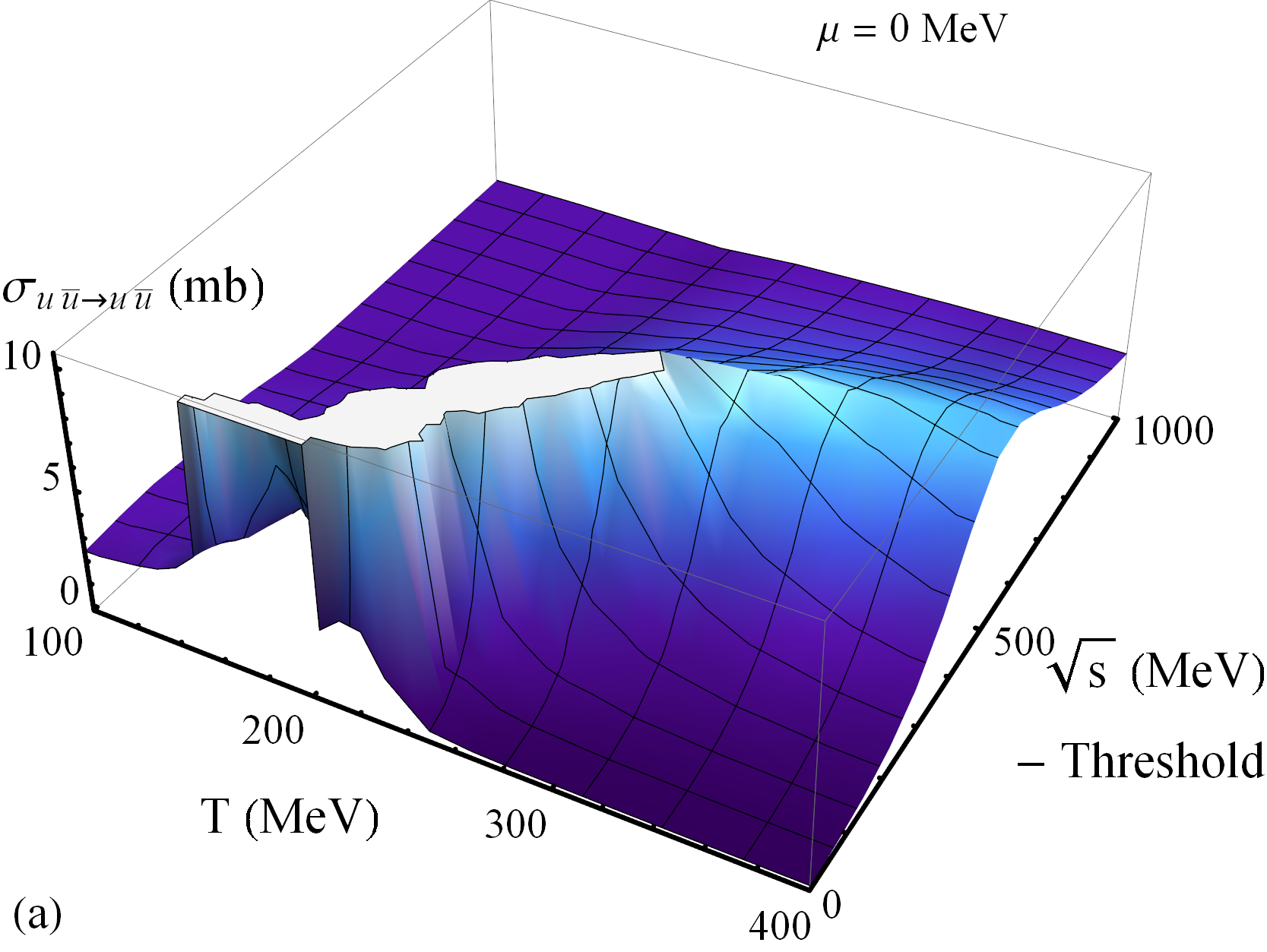} \ \ \
    \includegraphics[width=8.5cm]{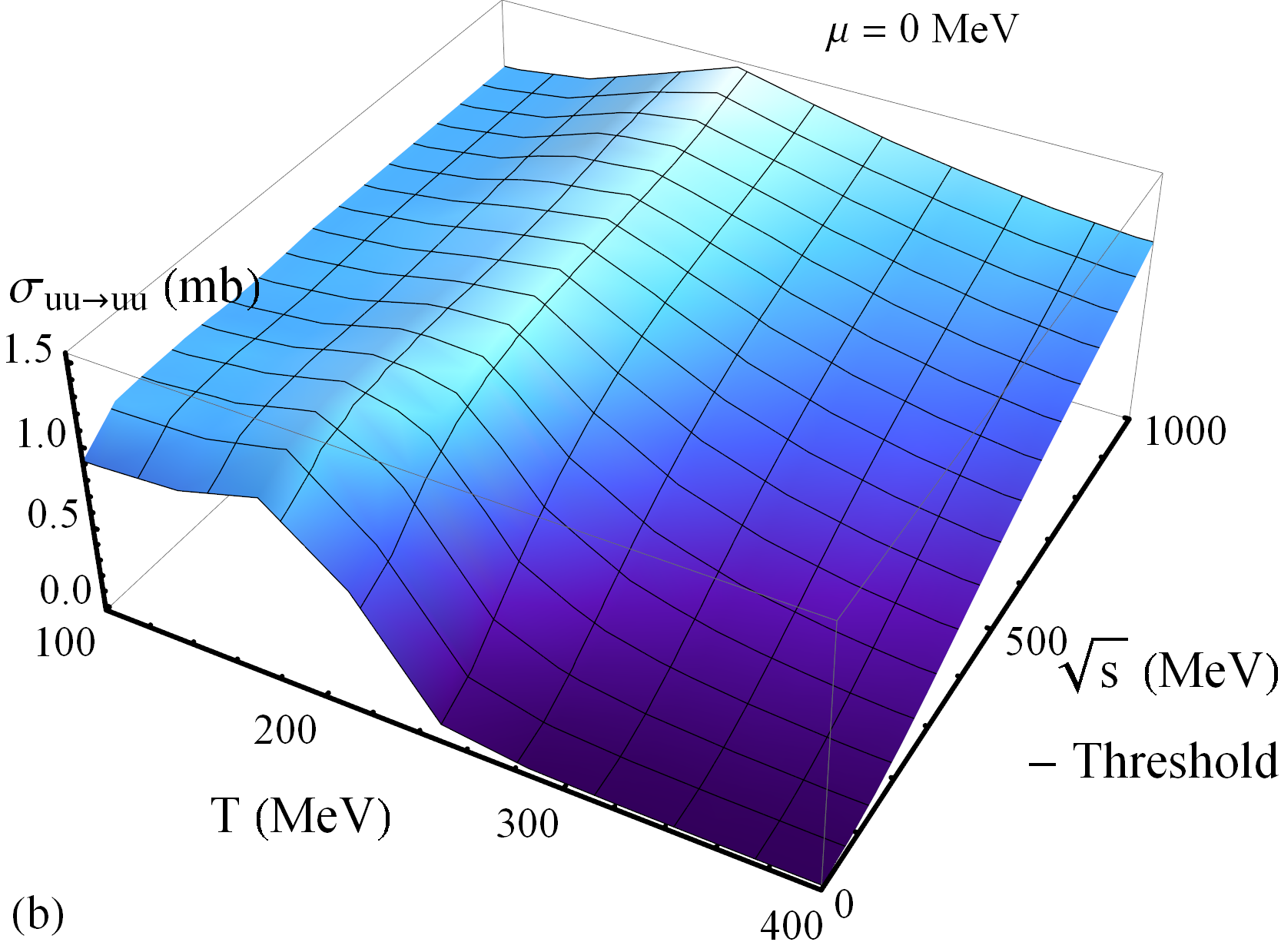}
  \end{center}
  \vskip -3mm
  \caption{(Color online) Cross sections $u\bar{u} \to u\bar{u}$ (a) and $uu \to uu$ (b) as a function of $(T,\sqrt{s}-\sqrt{s}_0)$ in the NJL model where $\sqrt{s}_0$ denotes the threshold.\label{CrossSection3D}}
\end{figure*}

The NJL model is a non renormalizable model which has to be regularized in the ultraviolet limit using a cutoff parameter $\Lambda$. The set of parameters for our calculations is displayed in Table \ref{paramNJL}.

We construct the meson propagator as a series of $q/\bar q$ polarization loops which give
\begin{equation}
  \frac{- i g^2_{\pi q \bar q}}{k^2 - M^2} = \frac{2 i G}{1 - 2 G \Pi(k)},
\end{equation}
with $\Pi(k)$ being the $q$/$\bar{q}$ polarization loop (see \cite{Rehberg1996a}). The mass $M$ of the meson is obtained by solving the equation
\begin{equation}
  1 - 2G \Pi(k)\big|_{k^2=M^2} = 0.
\end{equation}
For our study we have a preferential interest in the partonic degrees of freedom and therefore we refer the reader to Refs. \cite{Klevansky1992,Vogl1991} for further information about the basics on the NJL model.

Masses of quarks and pseudoscalar mesons from the NJL calculations are displayed in Fig. \ref{MassNJL}. At low temperatures $T$ the spontaneous breaking of chiral symmetry generates an effective mass for the quarks. In contrast, the pseudoscalar mesons show an increasing mass with increasing temperature. We can define a Mott temperature as that temperature when the mass of mesons is equal to the mass of the intrinsic quarks plus antiquarks, i.e., $M =  m_q + m_{\bar q}$.

It is also possible to compute these masses at finite chemical potential $\mu$ as shown in Fig. \ref{Mass3D}. For small temperature $T$ and large chemical potential $\mu$, the cross over transition between the hadronic and partonic phases turns to a first-order phase transition in the NJL model \cite{Nebauer2002,Buballa2005}.

The Lagrangian \eqref{NJL_lagrangian} allows us to compute cross sections using quark or meson exchange diagrams \cite{Thomere2009}. We can see in Fig. \ref{CrossSection3D}(a) that in the case of quark-antiquark scattering the resonant $s$ channel gives very large cross sections around the Mott temperature, whereas the cross section for quark-quark scattering [Fig. \ref{CrossSection3D}(b)] stays roughly flat and is small. Note that different mesons have different Mott temperatures. We define for the rest of this study a critical temperature $T_c = T_{\text{Mott}} = 200$ MeV which is about the mean value of the different Mott temperatures in the NJL model.

\section{The Polyakov--Nambu--Jona-Lasinio model}
\vskip -3mm
The NJL model assumes the gluon degrees of freedom to be integrated out and therefore confinement is not included in the model. However, gluons can be included on the level of a chiral-point coupling of quarks together with a static background field representing the Polyakov loop \cite{Fukushima2003}. This Polyakov-loop extended NJL model--denoted as PNJL--was developed by fitting lattice pure gauge data and considering additionally three flavors of quarks in Refs. \cite{Ratti2007,Costa2008,Costa2010}. The PNJL Lagrangian used in Ref. \cite{Costa2008} is
\begin{equation}
  \begin{aligned}
    \mathscr{L}_{PNJL} =& \bar{\psi} \left( i D\!\!\!\!\!\!/ - m_0 \right) \psi + \mathcal{U}(\phi,\bar\phi,T) + \mu\bar{\psi}\gamma_0\psi\\
             &+ \ G \sum_{a=0}^8 \left[   \left( \bar{\psi}            \lambda^a \psi \right)^2
                                         + \left( \bar{\psi} i \gamma_5 \lambda^a \psi \right)^2 \right] \\
             &+ \ K \left[   \mathrm{det} \bar{\psi} \left( 1 - \gamma_5 \right) \psi
                              + \mathrm{det} \bar{\psi} \left( 1 + \gamma_5 \right) \psi \right].
  \end{aligned}
  \label{PNJL_lagrangian}
  \vspace{2mm}
\end{equation}
\vskip 5mm
\noindent which differs from Eq. \eqref{NJL_lagrangian} by adding an external gauge field $D^\mu = \partial^\mu - iA^\mu$ with $A^\mu = \delta^\mu_0 A^0$ (Polyakov gauge), in Euclidean notation $A^0 = -iA_4$, and by adding an effective potential $\mathcal{U}$ which depends on the Polyakov loop $\phi$:
\begin{equation}
\Phi(\vec x) = \frac{1}{N_c}{\rm Tr}_c\left\langle \left\langle L(\vec x)\right\rangle\right\rangle,
\label{eq:phi}
\end{equation}
where
\begin{equation}
L(\vec x) ={\cal P}\exp\Bigg[i\int_0^\beta d\tau A_4(\vec x, \tau)\Bigg].
\label{eq:loop}
\end{equation}
One can also extend this approach to finite baryon densities $n_B$ by adding $\mu\bar{\psi}\gamma_0\psi \approx \mu \langle\langle \psi^\dagger \psi \rangle\rangle = - \mu n_B$ to the Lagrangian \eqref{PNJL_lagrangian}. Note, however, the Polyakov loop potential might vary additionally explicitly with $\mu$.

\begin{figure*}
  \begin{center}
    \includegraphics[width=8.5cm]{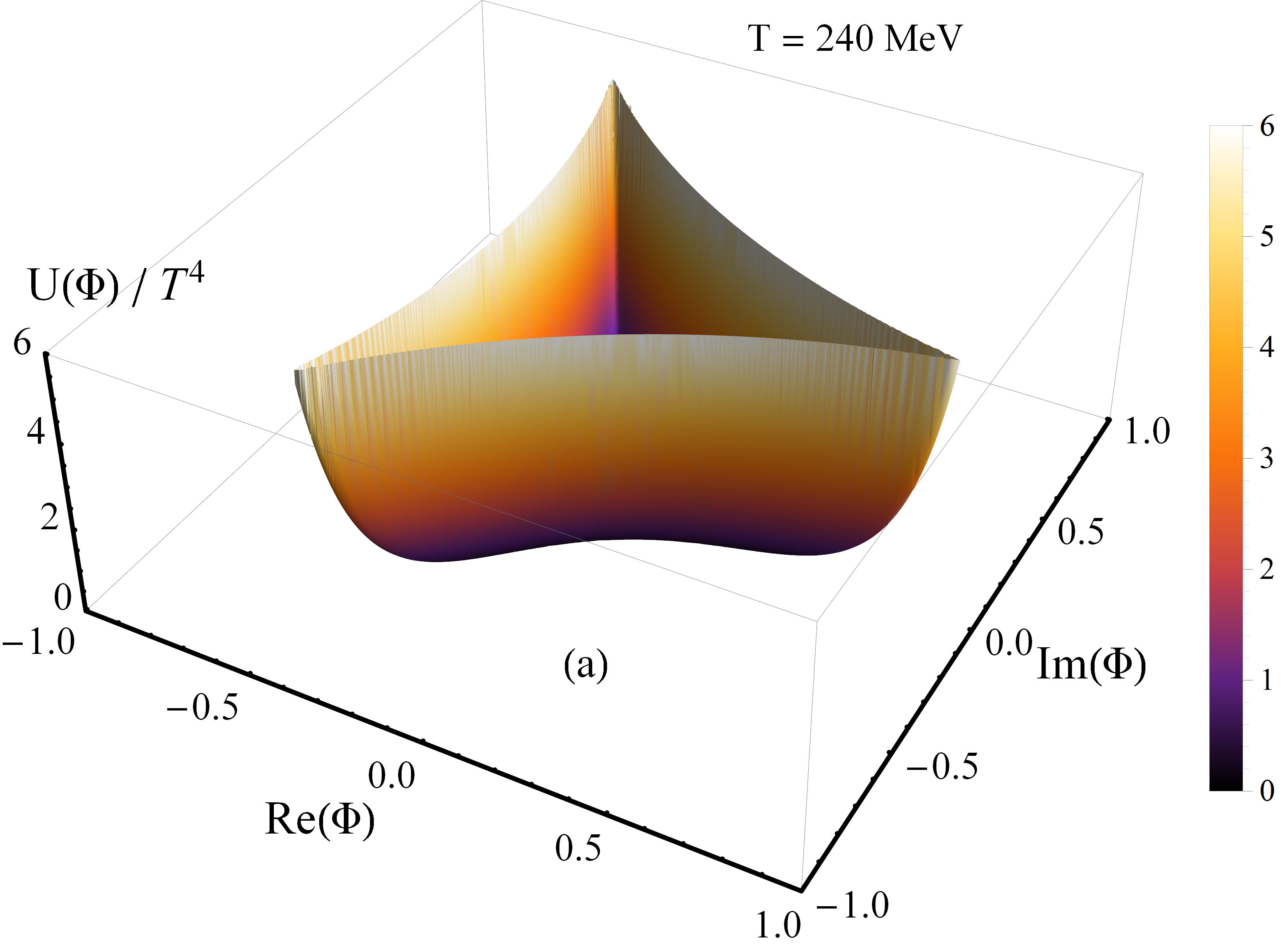} \ \ \
    \includegraphics[width=8.5cm]{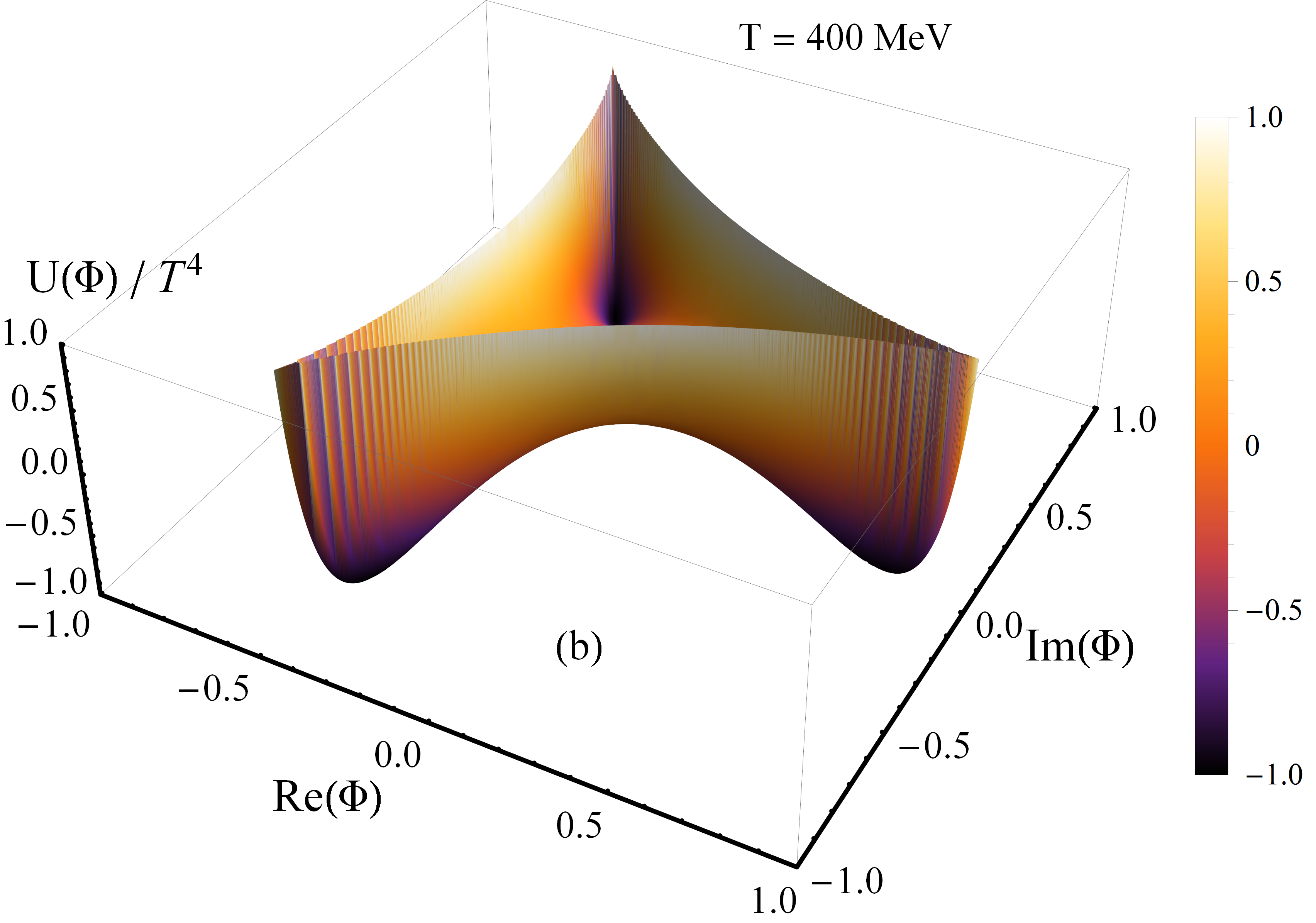}
  \end{center}
  \vskip -5mm
  \caption{(Color online) Polyakov potential $\mathcal{U}$ as a function of the real and imaginary part of the Polyakov loop $\phi$ for $T < T_c$ (a) and $T > T_c$ (b).\label{PolyakovPotential}}
\end{figure*}

\vskip 10mm
Linking the Polyakov loop to the quark degrees of freedom then modifies the Fermi-Dirac distribution to
\begin{equation}
  \begin{aligned}
  f_q \to&f_q^\Phi({\bf p},T,\mu)\\
  =&\frac{ ( \bar\Phi + 2\Phi e^{-(E_{\bf p} - \mu)/T} ) e^{-(E_{\bf p} - \mu)/T} + e^{-3(E_{\bf p} - \mu)/T} }
  {1 + 3( \bar\Phi + \Phi e^{-(E_{\bf p} - \mu)/T} ) e^{-(E_{\bf p} - \mu)/T} + e^{-3(E_{\bf p} - \mu)/T}},\\
  f_{\bar q} \to&f_{\bar q}^\Phi({\bf p},T,\mu)\\
  =&\frac{ ( \Phi + 2\bar\Phi e^{-(E_{\bf p} + \mu)/T} ) e^{-(E_{\bf p} + \mu)/T} + e^{-3(E_{\bf p} + \mu)/T} }
  {1 + 3( \Phi + \bar\Phi e^{-(E_{\bf p} + \mu)/T} ) e^{-(E_{\bf p} + \mu)/T} + e^{-3(E_{\bf p} + \mu)/T}}.
\label{fpnjl}
  \end{aligned}
\end{equation}

\begin{table} [!b]
  \caption{Parameters for the effective potential in the pure gauge sector [Eq.~\eqref{Ueff}] \cite{Ratti2007}.\label{paramPG}}
  \begin{ruledtabular}
    \begin{tabular}{ c c c c c }
      $a_0$ & $a_1$ & $a_2$ & $b_3$ & $T_0$ \\
      \hline
      \noalign{\vskip 3pt}
      3.51  & $-$2.47 &  15.2 & $-$1.75 &  270 MeV \\
    \end{tabular}
  \end{ruledtabular}
\end{table}

This distribution approaches the standard definition [Eq. \eqref{fermidis}] for $\Phi \to 1$ in the deconfined phase. For $\Phi \to 0$, the quarks are supposed to be confined in neutral  bound states and therefore the exponent in the Boltzmann  distribution increases by a factor of 3.

The static background field--first defined in Ref. \cite{Hansen2006}, and then modified in \cite{Friesen2011}--is assumed to be
\begin{equation}
  \begin{aligned}
\frac{\mathcal{U}(\Phi, \bar\Phi, T)}{T^4} =&-\frac{a(T)}{2} \bar\Phi \Phi + b(T) \times\\
&\ln[1 - 6 \bar\Phi \Phi + 4 (\bar\Phi^3 + \Phi^3) - 3 (\bar\Phi \Phi)^2 ],
\label{Ueff}
  \end{aligned}
\end{equation}
where
\begin{equation}
  \begin{aligned}
    a(T)&= a_0 + a_1 \left(\frac{T_0}{T}\right) +a_2 \left(\frac{T_0}{T}\right)^2,\\
    b(T)&= b_3 \left(\frac{T_0}{T}\right)^3.
  \end{aligned}
\end{equation}
From pure gauge lattice data \cite{Boyd1996,Kaczmarek2002} one can extract the parameters which are depicted in Table \ref{paramPG}.

The quark masses as well as of $\Phi$ and $\bar \Phi$ are obtained by minimizing the grand canonical potential \cite{Costa2008} for all these four degrees of freedom. For the masses one finds the same gap equation as Eq. \eqref{massNJL} for NJL. Only the Fermi-Dirac distributions are replaced by Eq. \eqref{fpnjl}:
\begin{equation}
  \langle\langle \bar{\psi}\psi \rangle\rangle
  = - 2 N_c \ \int_0^\Lambda \frac{d^3 p}{(2\pi)^3}
  \frac{m}{E_{\bf p}} [1 - f_q^\Phi - f_{\bar q}^\Phi].
  \label{PNJL_condensate}
\end{equation}

\begin{figure} [!b]
  \begin{center}
    \includegraphics[width=8.5cm]{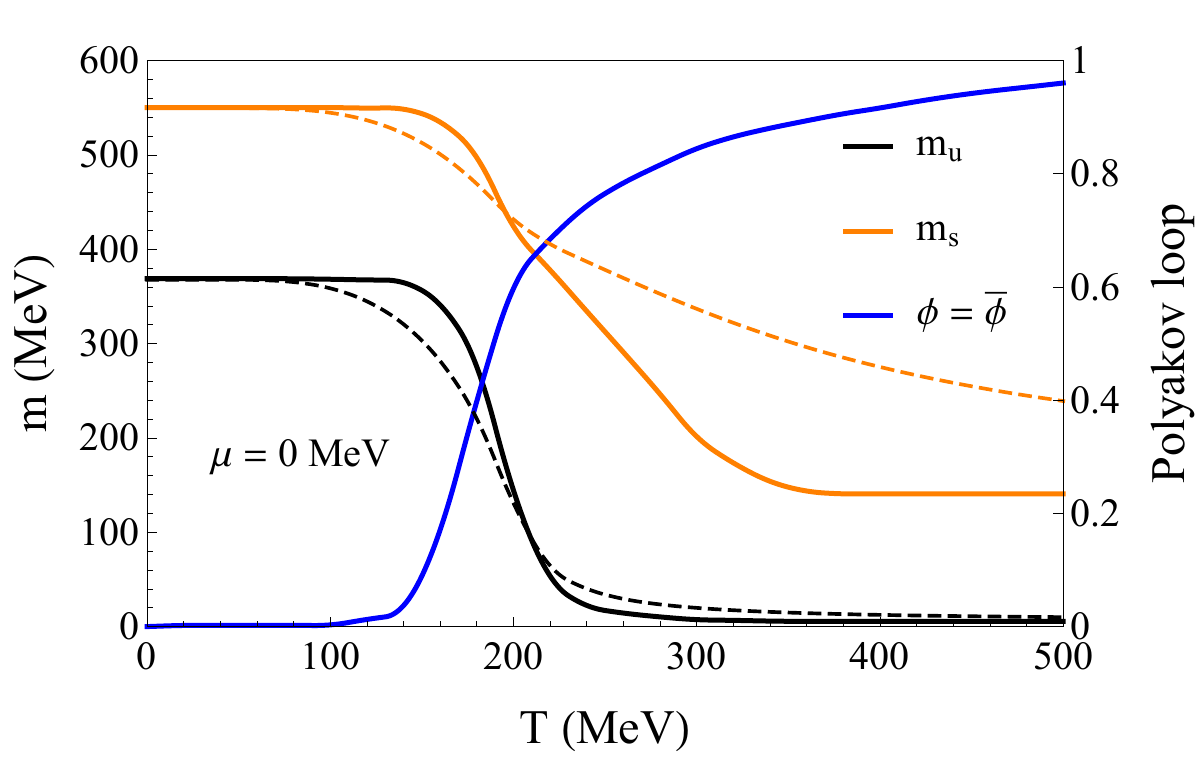}
  \end{center}
  \vskip -5mm
  \caption{(Color online) Masses of quarks and Polyakov loop $\Phi$ as a function of $T$ in the PNJL model (full lines) \cite{Costa2008} and in the NJL model (dashed lines).\label{MassPNJL}}
\end{figure}

The result for the masses and the Polyakov loop are displayed in Fig. \ref{MassPNJL}. For the comparison with the results of the NJL model  the gauge temperature $T_0$ has been modified to $T_0 = 210$ MeV. We see immediately the effect of the Polyakov loop on the phase transition: close to the transition the change of mass is faster for PNJL than NJL. Nevertheless, for $T \to 0$ or $\infty$ one recovers the same mass limit for the NJL and the PNJL models.
\vspace{-4mm}
\section{The dynamical quasi-particle model}
\vskip -3mm
The dynamical quasi-particle model \cite{Peshier2004,Peshier2005} describes QCD properties in terms of ``resummed'' single-particle Green's functions [in the sense of a two-particle irreducible (2PI) approach]. The parameters of the model are fitted in order to reproduce the equation of state from Ref. \cite{Borsanyi2010} above temperatures of 140 MeV.

This approach does not involve an effective Lagrangian since the strong coupling and propagators are assumed to represent ``resummed'' quantities. The 2PI framework then guarantees a consistent description of systems in and out of equilibrium on the basis of Kadanoff-Baym equations \cite{Kadanoff1962}. The resummed propagators are specified by complex self-energies that are temperature dependent. Whereas the real part of the self-energy describes a dynamically generated mass the imaginary part provides the interaction width of partons \cite{Bratkovskaya2011}; the mass and the width are linked in turn to the coupling constant $g^2(T)$ for strongly interacting matter, which is found to be enhanced in the infrared region (or low temperature).

The running coupling constant $g^2(T)$ for partons is approximated (for $T > T_c$, $\mu_q = 0$) by
\begin{equation}
  g^2(T/T_c) = \frac{48\pi^2}{(11N_c - 2N_f) \ln[\lambda^2(T/T_c - T_s/T_c)^2]},
  \label{runni}
\end{equation}
where the parameters $\lambda = 2.42$ and $T_s/T_c = 0.56$ have been extracted from a fit to the lattice data for purely gluonic systems ($N_f = 0$) \cite{Cassing2009a}. In Eq. \eqref{runni}, $N_c = 3$ stands for the number of colors, $T_c$ is the critical temperature ($= 158$ MeV), while $N_f$ denotes the number of flavors.

\begin{figure} [t]
  \begin{center}
    \includegraphics[width=8.5cm]{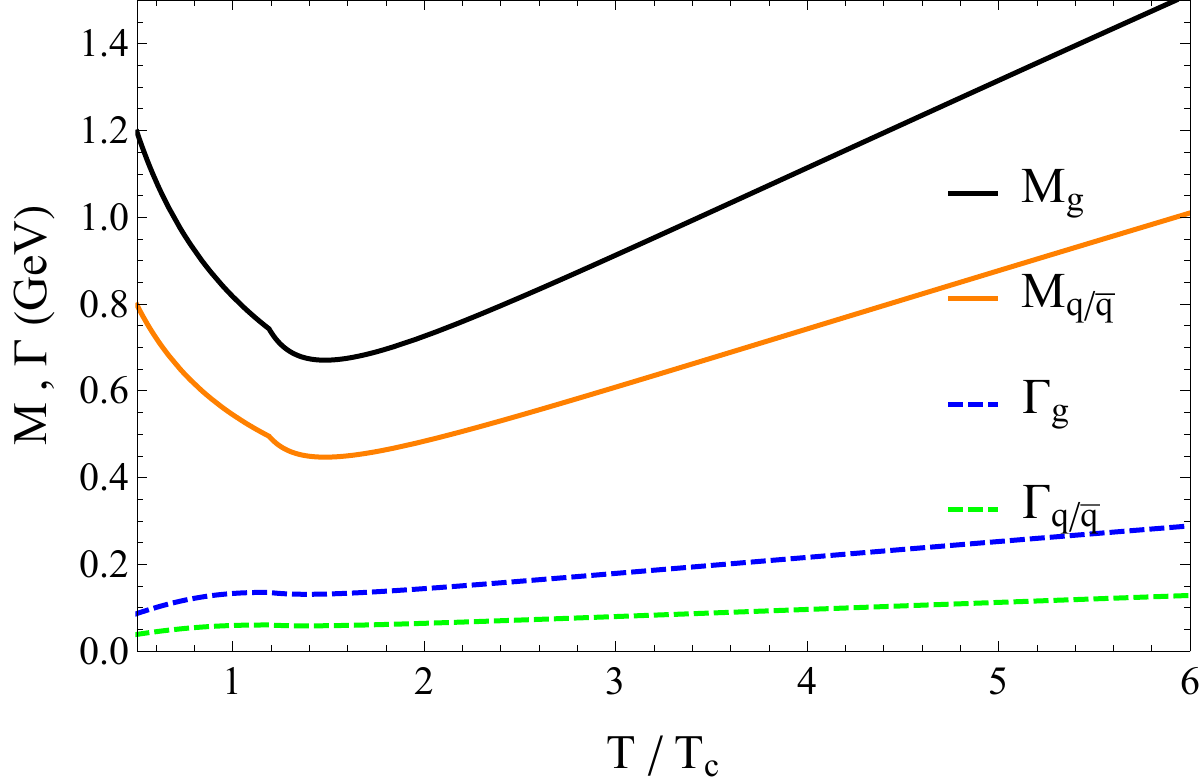}
  \end{center}
  \vskip -5mm
  \caption{(Color online) Masses and widths of quarks and gluons as a function of $T/T_c$ in the DQPM \cite{Ozvenchuk2012b}.\label{MassDQPM}}
  \vskip -5mm
\end{figure}

The functional forms for the dynamical gluon masses are chosen that they become identical to the perturbative thermal masses in the asymptotic high-temperature regime \cite{LeBellac1996}, i.e.,
\begin{equation}
  M^2_g(T,\mu_q) = \frac{g^2}{6} \left( \left( N_c + \frac{N_f}{2} \right) T^2 + \frac{3}{2} \sum_q \frac{\mu^{2}_q}{\pi^2} \right).
\end{equation}
The masses of quarks (antiquarks) are taken as
\begin{equation}
  M^2_{q/\bar q}(T,\mu_q) = \frac{N^2_c - 1}{8N_c} g^2 \left( T^2 + \sum_q \frac{\mu^2_q}{\pi^2} \right)
  \label{Mq9}
\end{equation}
following \cite{Peshier2002,Peshier2005b}; they differ from the hard-thermal-loop
(HTL) result \cite{LeBellac1996} by a factor of 2 and allow for a better description of the QCD equation of state from the lattice in the temperature range $0.6 T_c < T < 3 T_c$. In order to match the high-temperature HTL limit a smooth decrease of the additional factor $2 \rightarrow 1$ with temperature $T$ will have to be included in addition. Note that in the DQPM the coupling $g^2$ is assumed to be the running coupling $g^2(T)$ \eqref{runni}. The effective degrees of freedom (quarks, antiquarks, and gluons) in the DQPM have finite widths, which for $\mu_q = 0$ read
\setlength{\abovedisplayskip}{5px}
\setlength{\belowdisplayskip}{5px}
\begin{equation}
  \begin{aligned}
  \Gamma_g(T) =&\frac{1}{3} N_c \frac{g^2 T}{8\pi} \ln\left( \frac{2c}{g^2} + 1 \right),\\
  \Gamma_{q/\bar q}(T) =&\frac{1}{3} \frac{N^2_c - 1}{2N_c} \frac{g^2 T}{8\pi} \ln\left( \frac{2c}{g^2} + 1 \right),
  \end{aligned}
  \label{width}
\end{equation}
where the parameter $c = 14.4$ is related to a magnetic cut off.

To fit the equations of state of lattice QCD (from Ref. \cite{Borsanyi2010})  for three flavors, the coupling constant has been modified around $T_c$ and below and reads:
\begin{equation}
  g^2(T/T_c) \to g^2(T^\star/T_c) \left( \frac{T^\star}{T} \right)^{\gamma},
\end{equation}
where $T^\star = 1.19 T_c$ and $\gamma = 3.1$. The masses and widths used in the DQPM are displayed in Fig. \ref{MassDQPM} as a function of $T/T_c$.

Due to the finite imaginary parts of the self-energies, the parton spectral functions, i.e., the imaginary parts of the propagators, are no longer $\delta$ functions in the invariant mass squared but have a Breit-Wigner form
\begin{equation}
  \rho(\omega,{\bf p}) = \frac{\Gamma}{E} \left( \frac{1}{(\omega-E)^2 + \Gamma^{2}} - \frac{1}{(\omega+E)^2 + \Gamma^{2}} \right)
  \label{20}
\end{equation}
with the notation $E^2({\bf p}^2) = {\bf p}^2 + M^2 - \Gamma^2$, for quarks, antiquarks, and gluons. The spectral function (\ref{20}) is antisymmetric in $\omega$ and is normalized as
\begin{equation}
  \int_{-\infty}^{\infty}\frac{d\omega}{2\pi}\ \omega\rho(\omega,{\bf p})=
  \int_{0}^{\infty}\frac{d\omega}{2\pi}\ 2\omega \rho(\omega,{\bf p})=1.
  \label{norm}
\end{equation}
Equations (\ref{runni}) (\ref{norm}) specify the DQPM (and its few parameters); all further quantities discussed in this study then are fully defined and do not require any further assumptions. Although the free parameters of the DQPM are fitted to the LQCD equation of state in equilibrium, its results for the various transport coefficients (investigated in this article) are a direct consequence within the model without incorporating any additional parameters.
\vspace{-4mm}
\section{Equations of state}
\vskip -3mm
\begin{figure*}
  \begin{center}
    \includegraphics[width=8.5cm]{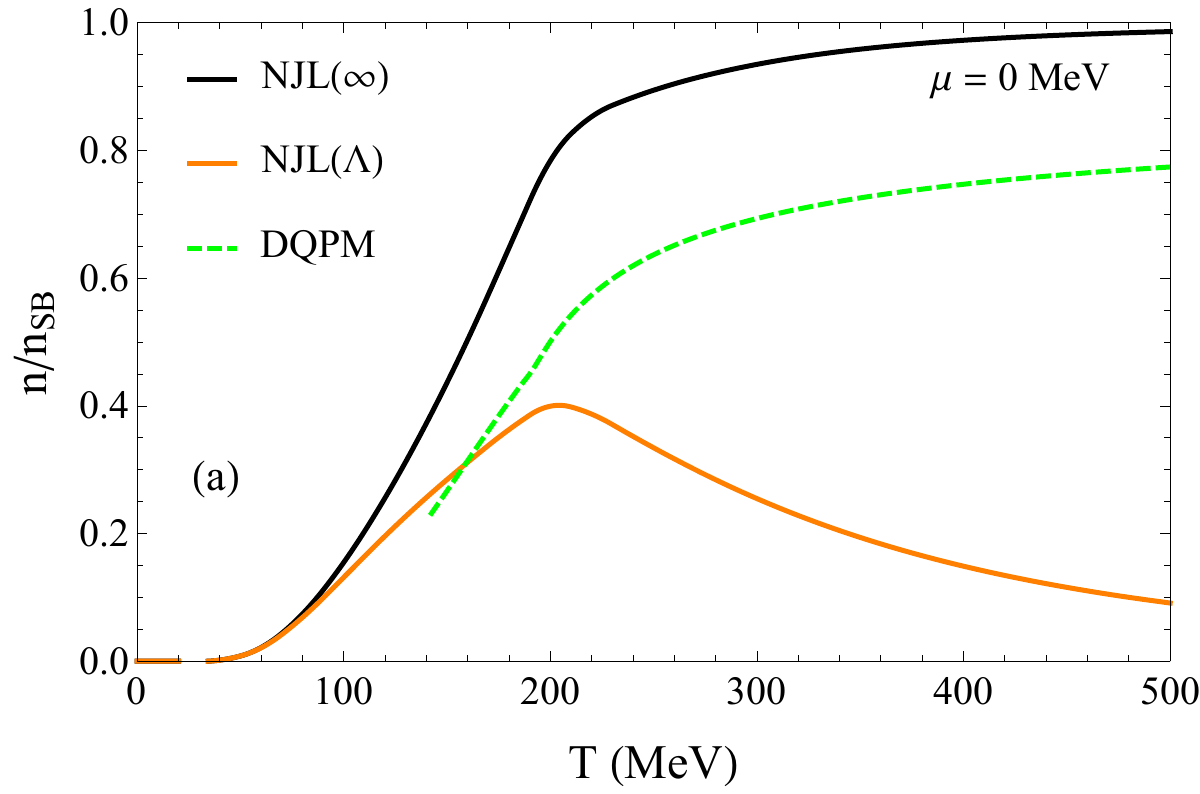} \ \ \
    \includegraphics[width=8.5cm]{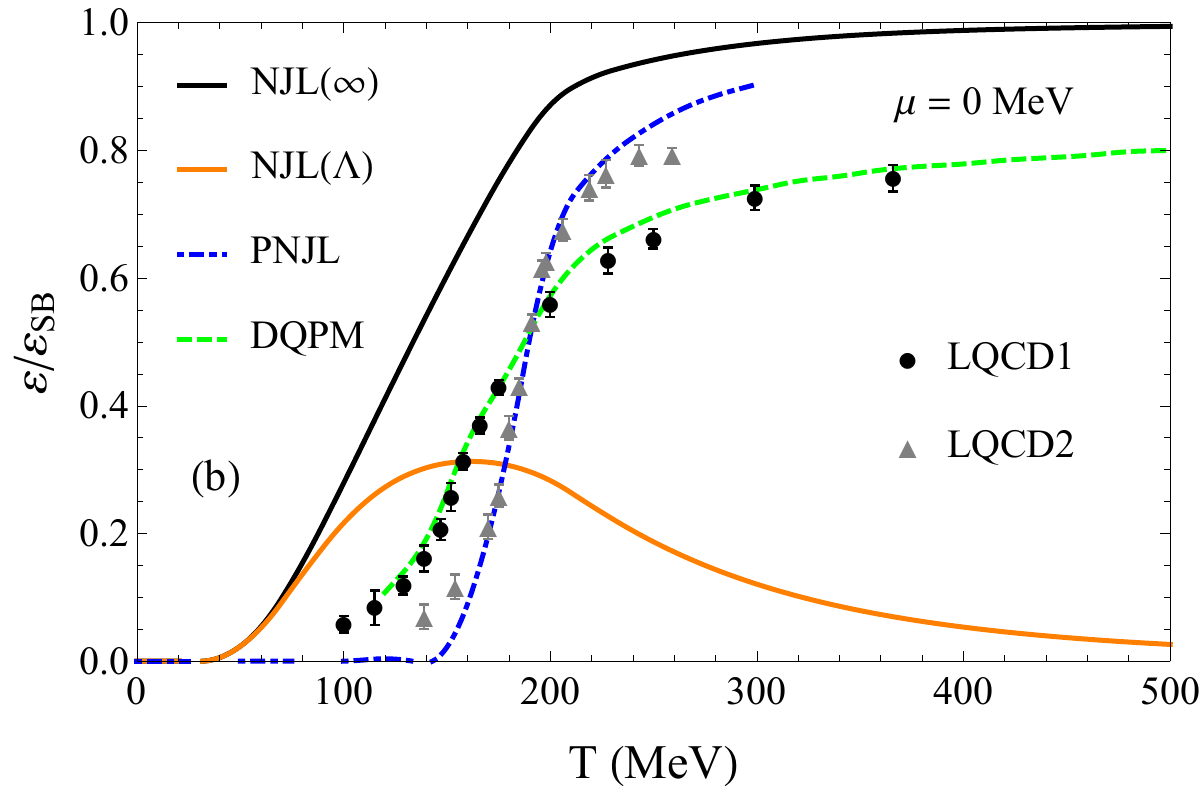}\\
    \includegraphics[width=8.5cm]{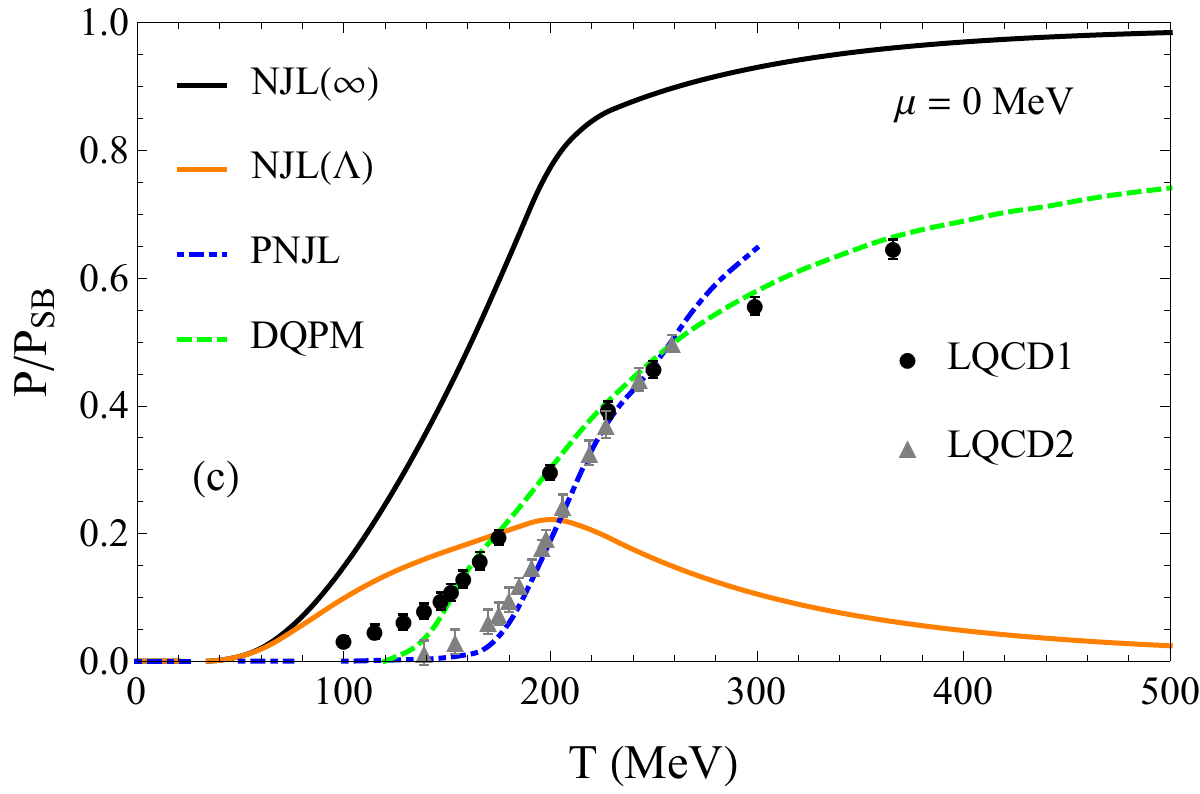} \ \ \
    \includegraphics[width=8.5cm]{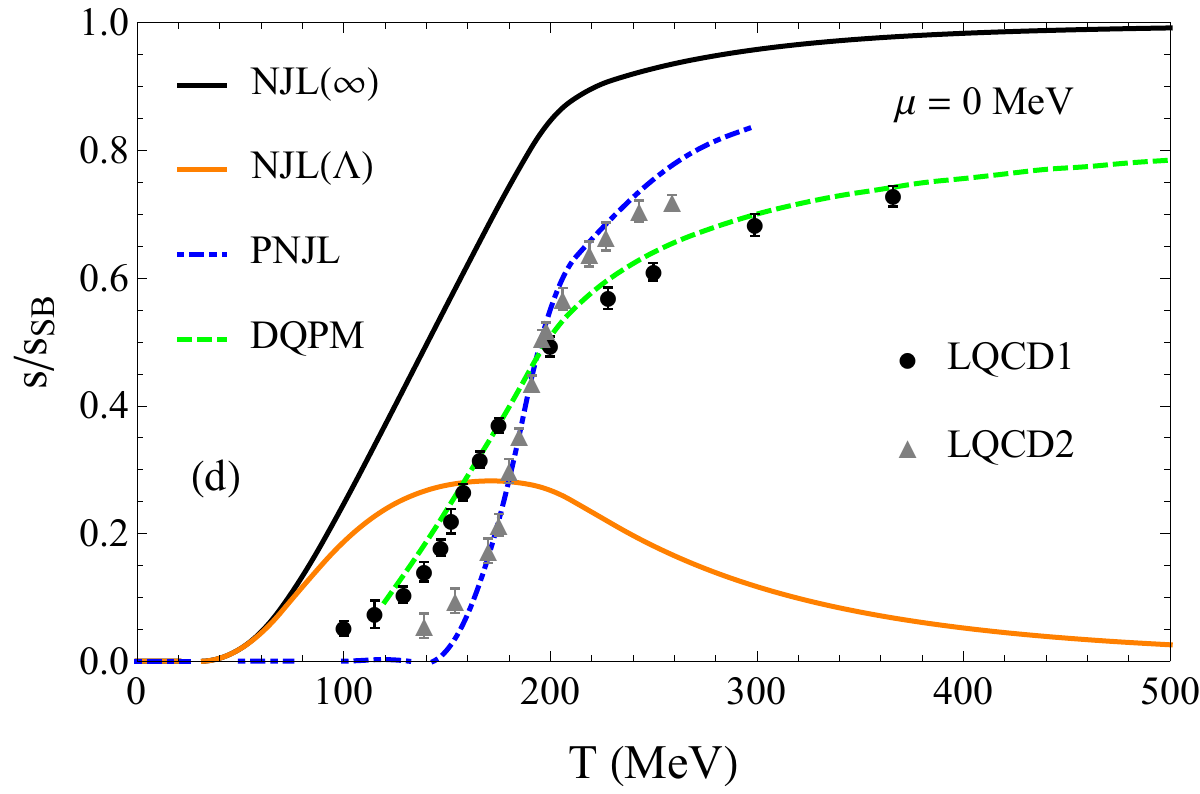}
  \end{center}
  \vskip -5mm
  \caption{(Color online) Particle density $n$ (a), energy density $\varepsilon$ (b), pressure density $P$ (c), and entropy density $s$ (d) as a function of temperature $T$ from different models compared to LQCD1 data from Ref. \cite{Borsanyi2010} and LQCD2 data from \cite{Cheng2010}. The PNJL results have been taken from Ref. \cite{Costa2010}. We note that the  NJL calculations of condensates (or mean fields) must be regularized using the cut off, but once the complete set of NJL masses is available at given temperature, these  masses are used for observables in integrals over the thermal distributions which have not been regularized.\label{EoS}}
  \vskip -5mm
\end{figure*}

In order to compute the equation of state of strongly interacting matter, we use thermal distribution functions for partons, i.e. Eq. \eqref{fermidis} for quarks in the  NJL and DQPM approach and Eq. \eqref{fpnjl} for quarks in the PNJL model. For the gluons in the DQPM we use the Bose-Einstein distribution,
\begin{equation}
    f_g({\bf p},T) = (\exp[E_{\bf p} / T] - 1)^{-1},
\end{equation}
with $E_{\bf p}({\bf p},T,\mu) = \sqrt{{\bf p}^2 + m_g^2}$ and $m_g(T,\mu)$ being an effective mass for gluons. We note that the temperature and density dependence of the different quark masses $m_q(T,\mu)$ are rather different in the DQPM as compared to the (P)NJL model.

\setlength{\abovedisplayskip}{7px}
\setlength{\belowdisplayskip}{7px}

We compute the particle density $n$, energy density $\epsilon$, pressure $P$, and entropy density $s$ (cf. Fig. \ref{EoS}) in the NJL model as a function of temperature $T$ using the relations based on the definition of the stress-energy tensor $T^{\mu\nu}$ for non interacting particles:
\begin{equation}
  \begin{aligned}
   n(T,\mu) =&g_g     \int_{0}^{\infty} \frac{d^3 p}{(2\pi)^3}
                                                f_g\\
     &+ \frac{g_q}{6} \int_{0}^{\infty} \frac{d^3 p}{(2\pi)^3}
      \Bigg[ \sum_q^{u,d,s} f_q + \sum_{\bar q}^{\bar u,\bar d,\bar s} f_{\bar q} \Bigg],
  \end{aligned}
  \label{n_density}
\end{equation}
\vspace{-2mm}
\begin{equation}
  \begin{aligned}
   P(T,\mu) =&g_g     \int_{0}^{\infty} \frac{d^3 p}{(2\pi)^3}
                                                f_g \frac{{\bf p}^2}{3 E_g}\\
     &+ \frac{g_q}{6} \int_{0}^{\infty} \frac{d^3 p}{(2\pi)^3}
      \Bigg[ \sum_q^{u,d,s} f_q + \sum_{\bar q}^{\bar u,\bar d,\bar s} f_{\bar q} \Bigg] \frac{{\bf p}^2}{3 E_q},
  \end{aligned}
  \label{p_density}
\end{equation}
\vspace{-2mm}
\begin{equation}
  \begin{aligned}
\varepsilon(T,\mu) =&g_g \int_{0}^{\infty} \frac{d^3 p}{(2\pi)^3}
                                                f_g E_g\\
     &+ \frac{g_q}{6} \int_{0}^{\infty} \frac{d^3 p}{(2\pi)^3}
      \Bigg[ \sum_q^{u,d,s} f_q + \sum_{\bar q}^{\bar u,\bar d,\bar s} f_{\bar q} \Bigg] E_q,
  \end{aligned}
  \label{e_density}
\end{equation}
\vspace{-5mm}
\begin{equation}
  s(T,\mu) = \frac{\varepsilon(T,\mu) + P(T,\mu) - \mu n_B(T,\mu)}{T},
  \label{s_density}
\end{equation}
with the baryonic density
\begin{equation}
  n_B(T,\mu) = \frac{g_q}{6} \int_{0}^{\infty} \frac{d^3 p}{(2\pi)^3}
  \Bigg[ \sum_q^{u,d,s} f_q - \sum_{\bar q}^{\bar u,\bar d,\bar s} f_{\bar q} \Bigg],
\end{equation}
and the degeneracy factors
\begin{equation}
  \begin{aligned}
    &g_g = \text{polarization }[0,1]\times(N_c^2-1) = 16,\\
    &g_q = \text{spin }[+\text{\textonehalf},-\text{\textonehalf}]\times\text{parity }[q,\bar q]\times N_c N_f = 36.
  \end{aligned}
\end{equation}

The case of the (P)NJL model is treated differently, first of all because there are no gluons in the model but also because the model, is an effective approach and therefore it involves an ultraviolet cut off $\Lambda$ on the three-momentum to take into account the vacuum energy density of the quark condensate $\langle\langle \bar q q \rangle\rangle$ \cite{Klevansky1992}. When keeping this cut off in the calculation of the equations of state, we cannot reach the Stefan-Boltzmann limit (see Fig. \ref{EoS}, orange lines). For the following results we do not use this limit, considering a realistic medium where the momenta of the partons are thermally distributed and the cut off is removed in integrals of observables with thermal distributions.

The Stefan-Boltzmann limit is defined by
\begin{equation}
  n_{SB}(T) = \left( g_g + \frac{3}{4} g_q \right) \frac{\zeta(3)}{\pi^2}T^3
\end{equation}
with $\zeta(3) \simeq 1.202$ for the particle density and
\begin{equation}
  \begin{aligned}
  P_{SB}(T,\mu) =&\frac{g_g}{72} 0.8 \pi^2 T^4 +\\
                 &\frac{g_q}{72} \left( 0.7 \pi^2 T^4 + 3 T^2 \mu_q^2 + 1.5 \frac{\mu_q^4}{\pi^2} \right)
  \end{aligned}
\end{equation}
for the pressure, in which the first term stands for the gluon degrees of freedom and the second for that of the quarks. In the SB limit we have the relations
\begin{equation}
  \begin{aligned}
    \varepsilon_{SB}(T,\mu) =&3 P_{SB}(T,\mu),\\
              s_{SB}(T,\mu) =&4 P_{SB}(T,\mu) T.
  \end{aligned}
\end{equation}
In Fig. \ref{EoS} the SB limit is different for the NJL model and for PNJL/LQCD due to the lack of the gluon degrees of freedom. We recall that the results of the Polyakov-NJL model differ from the NJL results due to the explicit gluon potential $\mathcal{U}$ \cite{Costa2010}.

The critical temperature $T_c$ is also different between the NJL, PNJL, and LQCD. In the NJL model the definition of $T_c$ from the equation of state ($T_c \approx 170 MeV$) is different from the Mott temperatures for the different processes (see above). For convenience we choose an intermediate value $T_c = 200$ MeV. For the DQPM this value is taken from the Wuppertal-Budapest LQCD Collaboration \cite{Aoki2006,Borsanyi2010} as $T_c \approx 158$ MeV, and for the PNJL from the HotQCD LQCD Collaboration \cite{Cheng2010} as $T_c \approx 185$ MeV (a lattice value which is below the PNJL Mott temperatures). Note that the different critical temperatures from the LQCD collaborations are due to different discretizations of the action and different fermion masses, respectively.

The approach for calculating the equation of state in the DQPM is different but thermodynamically consistent. One starts from the evaluation of the entropy density $s$ within the quasiparticle approach \cite{Peshier2004,Cassing2008} which depends on the complex propagators and complex self-energies of the degrees of freedom as well as thermal Bose or Fermi distributions. Note that this entropy stands for an interacting system. Then using the thermodynamic relation
\begin{equation}
  s(T) = \left( \frac{\partial P(T)}{\partial T} \right)_\mu
\end{equation}
(for a fixed chemical potential $\mu$) one obtains the pressure $P$ by integration of $s(T)$ over $T$ while the energy density $\varepsilon$ is gained using Eq. \eqref{s_density}. A self-consistent check is done to compare the result with Eq. \eqref{e_density}. In the DQPM one has to include the off-shellness of the degrees of freedom by replacing
\begin{equation}
  f(E_{\bf p}) \to \int_0^\infty \frac{d\omega}{2\pi} f(\omega) \rho(\omega,{\bf p}) \ 2 \omega
\end{equation}
with the spectral function $\rho(\omega)$ [Eq. \eqref{20}] and then integrating over $\omega$ additionally.

\begin{figure} [!t]
  \begin{center}
    \includegraphics[width=8.5cm]{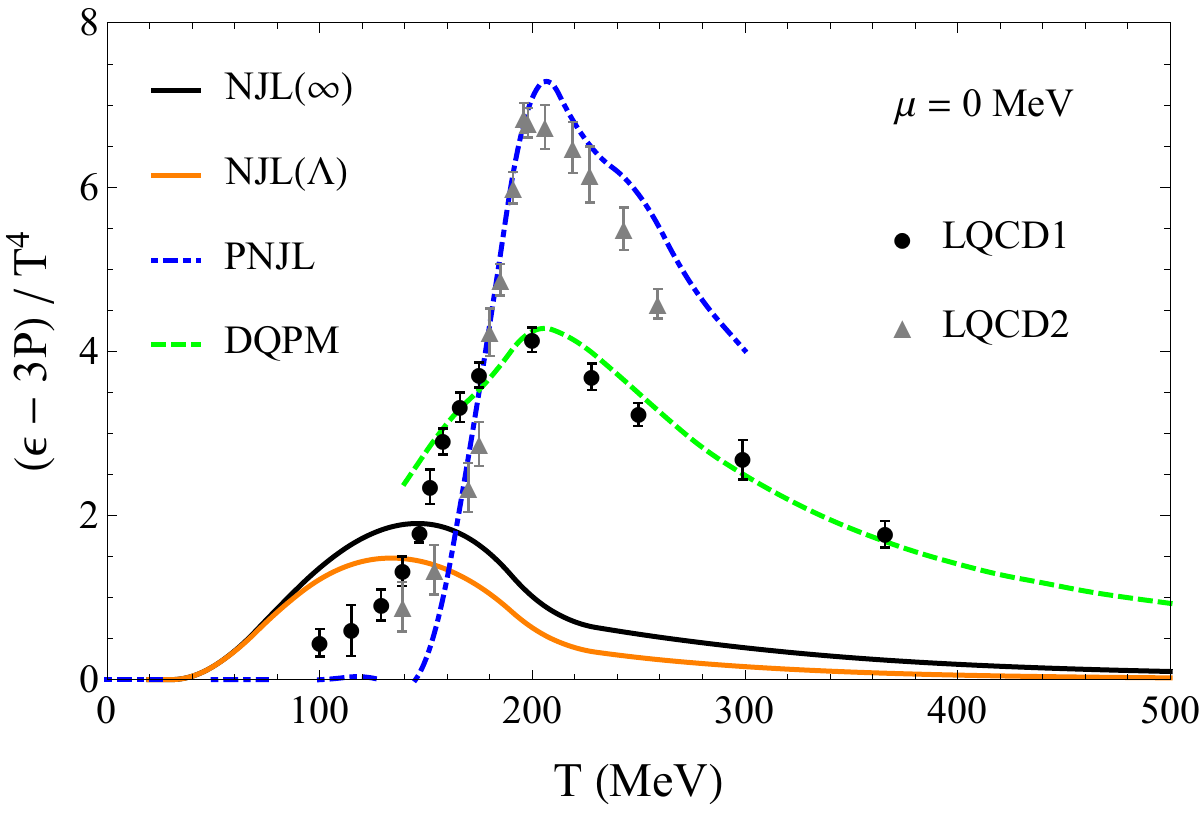}
  \end{center}
  \vskip -5mm
  \caption{(Color online) Trace anomaly $(\varepsilon - 3 P) / T^4$ as a function of $T$ from different models compared to LQCD data 1 from Ref. \cite{Borsanyi2010} and LQCD data 2 from \cite{Cheng2010}. The PNJL results have been taken from Ref. \cite{Costa2010}.\label{TraceAnomaly1}}
  \vskip -4mm
\end{figure}

We also note that the grand canonical potential $\Omega$ is used in the (P)NJL model to compute the pressure using the relation $P(T,\mu) = -\Omega(T,\mu)$.

In Fig. \ref{TraceAnomaly1} we display the interaction measure--known in LQCD as the trace anomaly--in comparison with LQCD and find again the NJL model to be in poor agreement with lattice data from Ref. \cite{Borsanyi2010,Cheng2010}. Note that the PNJL \cite{Costa2010} and DQPM calculations agree well with the LQCD data; however, both have been adjusted to different LQCD results, i.e. either to those from  Ref. \cite{Borsanyi2010} (DQPM) or from Ref. \cite{Cheng2010} (PNJL). We recall that the gluon pressure is taken into account in the PNJL model \cite{Ratti2007,Costa2010}.

The derivatives of the previous densities provide further interesting quantities. For instance we can compute the specific heat $c_V$ and the speed of sound squared $c_s^2$ as
\begin{equation}
  c_V = T \left( \frac{\partial s}{\partial T} \right)_V = T \left( \chi_{TT} - \frac{\chi_{\mu T}^2}{\chi_{\mu\mu}} \right)
\end{equation}
and
\begin{equation}
  c_s^2 = \left( \frac{\partial P}{\partial \varepsilon} \right)_{n_B} = \frac{s \chi_{\mu\mu} - n_B \chi_{\mu T}}{c_V \chi_{\mu\mu}}
\end{equation}
with the susceptibilities $\chi_{xy} = \partial^2 P / \partial x \partial y$.

\begin{figure} [!t]
  \begin{center}
    \includegraphics[width=8.5cm]{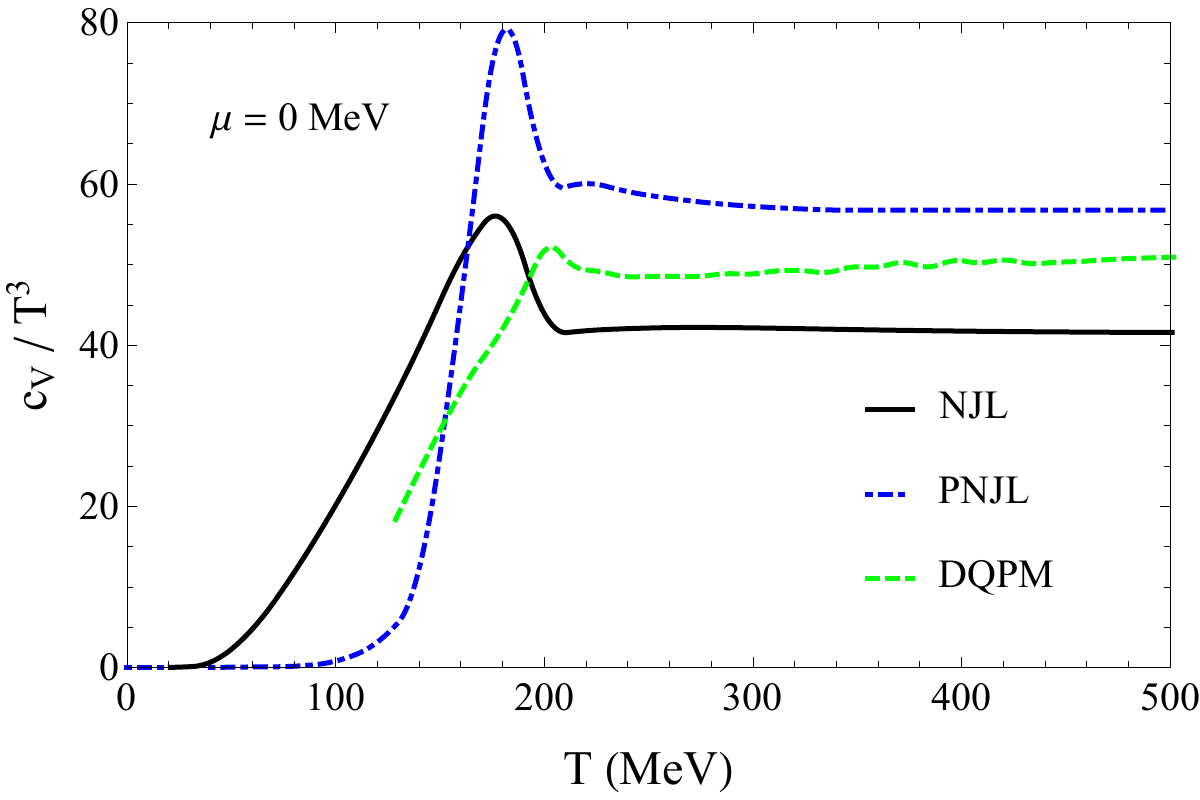}
  \end{center}
  \vskip -5mm
  \caption{(Color online) Specific heat $c_V$ as a function of $T$ in the NJL model, the DQPM and the PNJL model (taken from Ref. \cite{Bhattacharyya2013}).\label{SpecificHeat}}
\end{figure}

\begin{figure} [!b]
  \begin{center}
    \includegraphics[width=8.5cm]{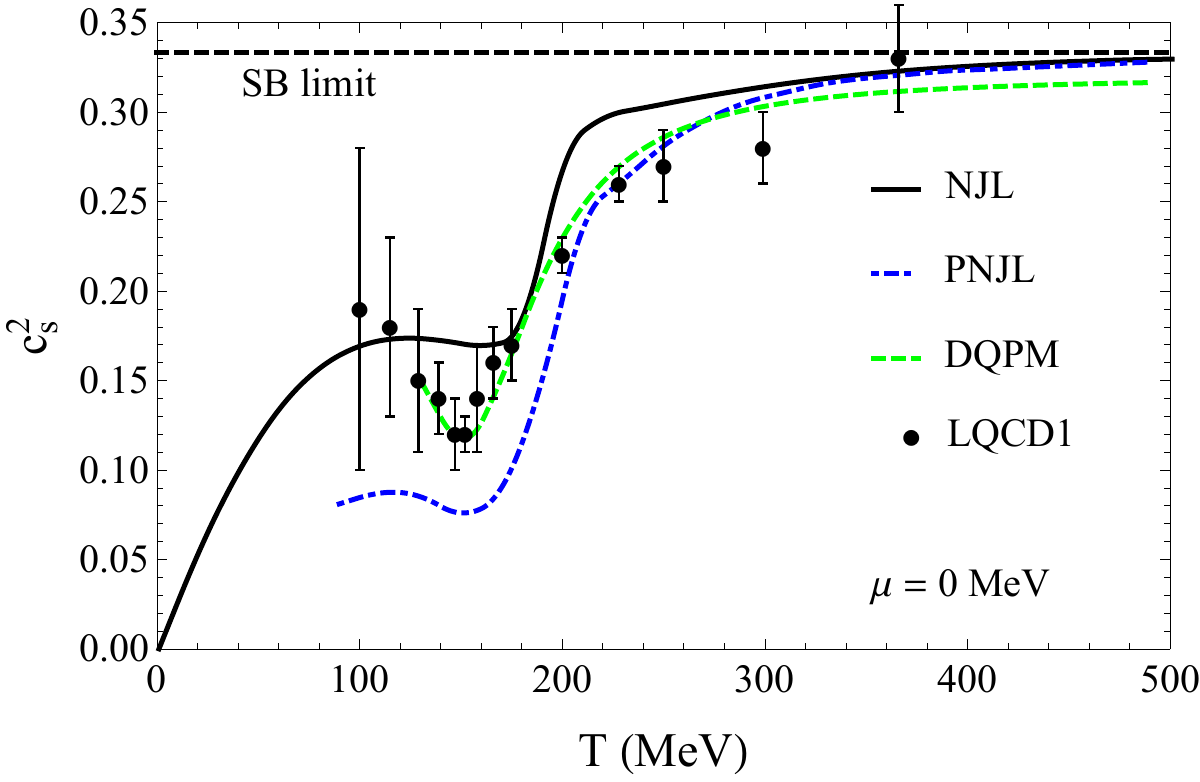}
  \end{center}
  \vskip -5mm
  \caption{(Color online) Speed of sound squared $c_s^2$ for the NJL model, for the PNJL model from Ref. \cite{Bhattacharyya2013}, and for the DQPM as a function of $T$ compared to LQCD data from Ref. \cite{Borsanyi2010}.\label{SpeedofSound}}
  \vskip -3mm
\end{figure}

Figure \ref{SpecificHeat} shows the scaled specific heat $c_v/T^3$ which reaches a limit above $T_c$. This limit is not the same for the different model calculations because the PNJL model and the DQPM include gluon degrees of freedom in the pressure, which changes the SB limit and the specific heat accordingly. Note that the parametrization of the PNJL model from Ref. \cite{Bhattacharyya2013} is fitted to different LQCD data than that from Ref. \cite{Costa2008}.

In Fig. \ref{SpeedofSound} we display the speed of sound squared $c_s^2(T)$ (from Ref. \cite{Borsanyi2010}) as a function of $T$. $c_s^2$ passes through a local minimum around the critical temperature and then reaches the SB limit ($=1/3$) at high temperature $T$. This minimum indicates a fast change in the masses of partons in the (P)NJL model (e.g. quarks) as well as in the DQPM (cf. Fig. \ref{MassDQPM}).

\section{Integrated cross sections}
\vskip -2mm
\begin{figure}
  \begin{center}
    \includegraphics[width=8.5cm]{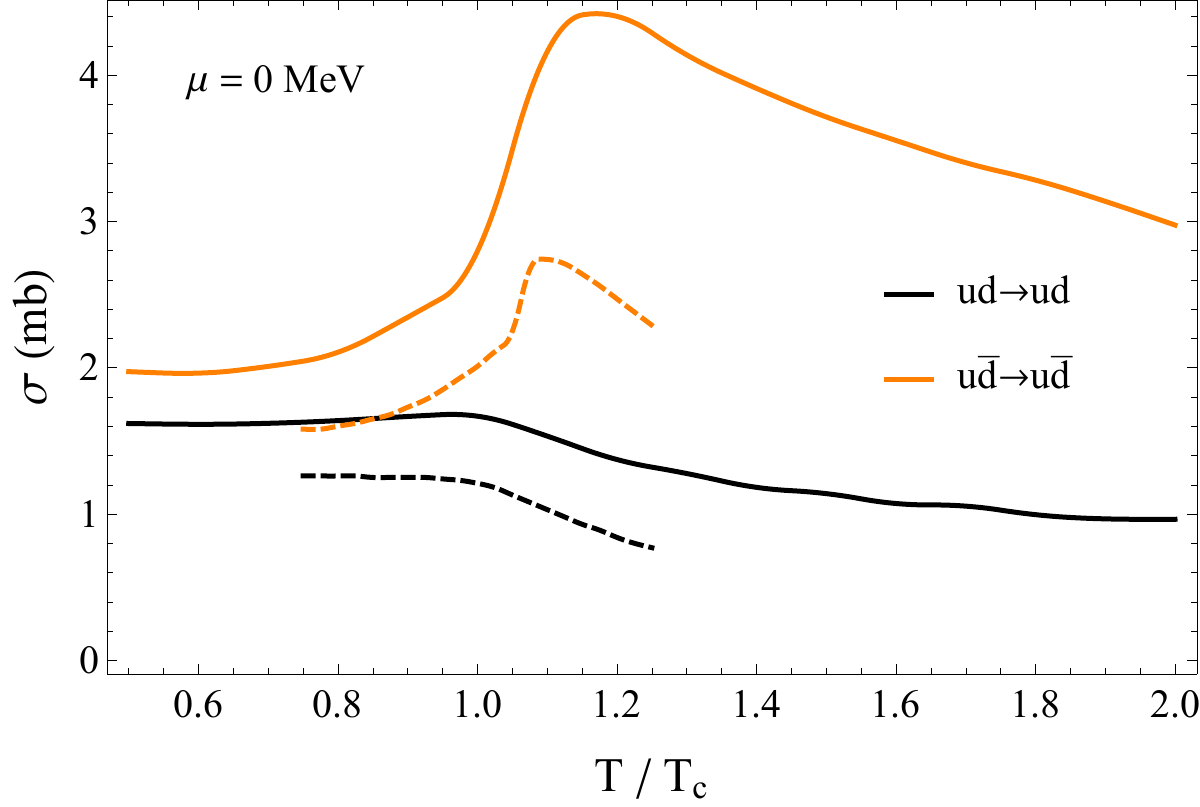}
  \end{center}
  \vskip -5mm
  \caption{(Color online) Cross sections as a function of $T$ in the NJL model (dashed lines from Ref. \cite{Rehberg1996b}).\label{IntegratedCrossSection}}
  \vskip -3mm
\end{figure}
\vskip -2mm
The first step before computing transport coefficients in the NJL model is to integrate the elastic cross sections $\sigma(T,\mu,s)$ over the invariant energy squared $s$ taking into account all possible kinematic reactions for a given thermodynamical medium in equilibrium for fixed $(T,\mu)$. The integration is done using \cite{Sasaki2010}
\begin{equation}
  \sigma(T,\mu) = \int_{\text{Th}}^\infty d s \ \sigma(T,\mu,s) \ L(T,\mu,s),
\end{equation}
with the threshold $\text{Th} =$ max($(m_1 + m_2)^2$,$(m_3 + m_4)^2$) [which depends on $(T,\mu)$] and $L$ being the probability of yielding a $q \bar q$ ($qq$) pair with the energy $\sqrt{s}$ in the medium $(T,\mu)$:
\begin{equation}
  \begin{aligned}
  L(T,\mu,s) =&C(T,\mu) \ E_1 E_2 \ p_{\text{cm}}(s) \ v_{\text{rel}}(s)\\
              &\times f_q\left(E_1 - \mu\right) f_{\bar q(q)}\left(E_2 \pm \mu\right)
  \end{aligned}
  \label{flux}
\end{equation}
with the center-of-mass momentum
\begin{equation}
   p_{\text{cm}}(s) = \frac{\sqrt{\left(s-(m_1+m_2)^2\right) \left(s-(m_1-m_2)^2\right)}}{2 \sqrt{s}},
\end{equation}
the relative velocity (in the center of mass)
\begin{equation}
  \begin{aligned}
   v_{\text{rel}}(s)
   =&\frac{\sqrt{(p_1 p_2)^2 - (m_1 m_2)^2}}{E_1 E_2}\\
   =&\frac{p_{\text{cm}}(s)\sqrt{s}}{E_1 E_2} = \frac{p_{\text{cm}}}{E_1} + \frac{p_{\text{cm}}}{E_2},
  \end{aligned}
  \label{vrel}
\end{equation}
while $C$ is a normalization factor fixed by
\begin{equation}
    C^{-1}(T,\mu) = \int_{\text{Th}}^\infty d s \ L(T,\mu,s).
\end{equation}
Similar calculations can be performed for cross sections with massive gluons by including the Bose-Einstein distribution in Eq. (\ref{flux}).

Now we can integrate the 11 different elastic NJL cross sections \cite{Rehberg1996b} over $s$:
\begin{itemize}
  \setlength{\itemsep}{0pt}
  \item[(i)] 4 $qq \to qq$ cross sections,
  \item[(ii)] 4 $q\bar q \to q\bar q$ cross sections,
  \item[(iii)] 3 $q\bar q \to q'\bar q'$ cross sections.
\end{itemize}
We show the results for the channels $ud \to ud$ and $u\bar d \to u\bar d$ in Fig. \ref{IntegratedCrossSection} which are in good qualitative agreement with the previous calculations from Ref. \cite{Rehberg1996b}. Nevertheless we note a small difference between the results which is due to a different normalization of the cross section in Ref. \cite{Rehberg1996b} as compared to ours (in line with Refs. \cite{Zhuang1995,Sasaki2010}). We can see that while the shape of the cross section is similar (e.g., flat $qq$ and peaked $q\bar q$ cross sections) the maximum is different.

For the DQPM we do not need the explicit cross sections since the inherent quasiparticle width $\Gamma(T)$ directly provides the total interaction rate.
\vspace{-4mm}
\section{Relaxation times}
\vskip -2mm
By means of the integrated cross sections we can compute the relaxation time $\tau$ of particles within the NJL model. This quantity relates to the mean-free-path $\ell$ between two elastic collisions as \cite{Sasaki2010}
\begin{equation}
  \tau^{-1}_i(T,\mu) = \sum_q n_q(T,\mu) \ \sigma_{iq}(T,\mu) \ = (v_{\text{rel}} \ell)^{-1},
\end{equation}
with $q = u,d,s,\bar u,\bar d,\bar s$, and the densities
\begin{equation}
  n_q(T,\mu) = \int_0^\infty \frac{d^3 p}{(2\pi)^3} f_q,\\
\end{equation}
In the DQPM approach one can directly use the flavor-blind reaction rates based on the particle width \eqref{width} \cite{Ozvenchuk2012b}:
\begin{equation}
  \tau^{-1}_q(T) = \Gamma_q(T)
  \text{ and }
  \tau^{-1}_g(T) = \Gamma_g(T).
\end{equation}

\begin{figure} [!t]
  \begin{center}
    \includegraphics[width=8.5cm]{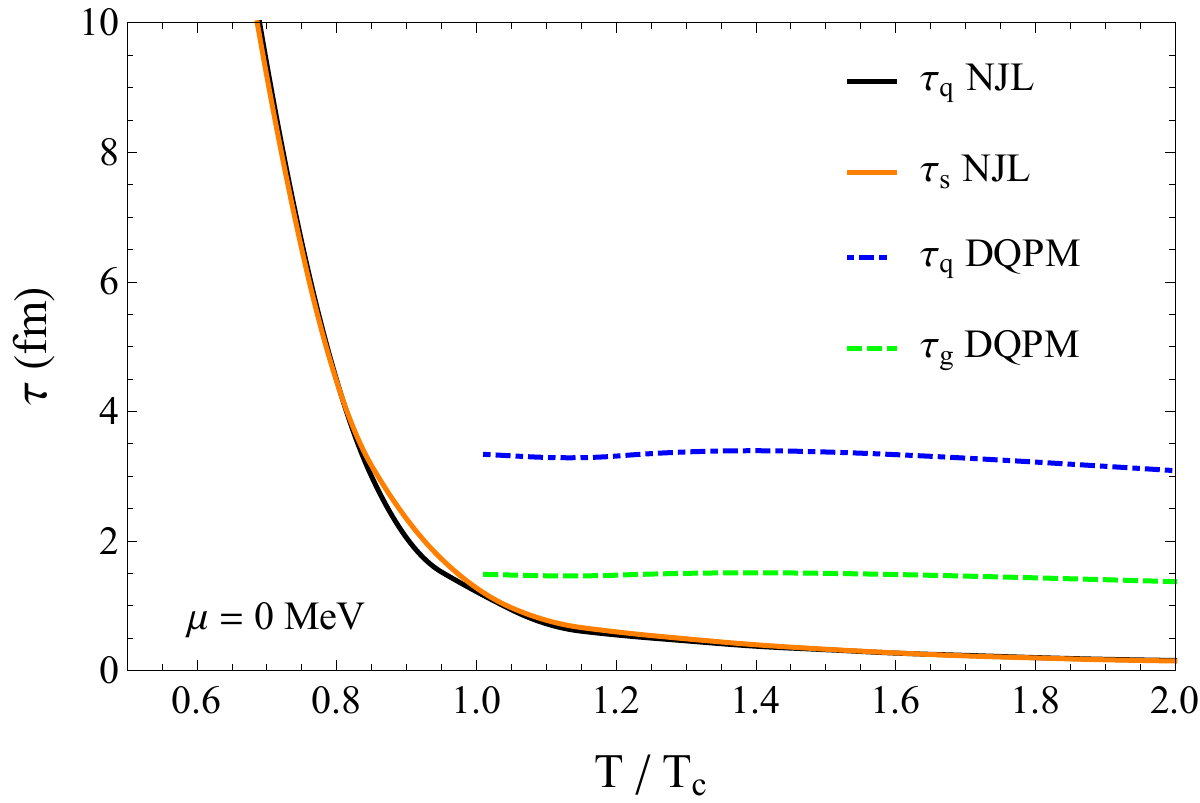}
  \end{center}
  \vskip -5mm
  \caption{(Color online) Relaxation times $\tau$ as a function of $T/T_c$ from the NJL model and the DQPM.\label{RelaxationTime}}
  \vskip -3mm
\end{figure}

For example the relaxation time in the NJL model for a $u$ quark (knowing that we have isospin symmetry for $u$ and $d$ quarks and no gluons) reads:
\begin{equation}
  \begin{aligned}
    \tau^{-1}_u      &=   n_u        \ (\sigma_{uu \to uu} + \sigma_{ud \to ud})
                        + n_{\bar u} \ (  \sigma_{u\bar u \to u\bar u}\\
                     &   \quad          + \sigma_{u\bar d \to u\bar d}
                                        + \sigma_{u\bar u \to d\bar d}
                                        + \sigma_{u\bar u \to s\bar s} ) \\
                     &   \quad + n_s \ \sigma_{us \to us}
                               + n_{\bar s} \ \sigma_{u\bar s \to u\bar s}.
  \end{aligned}
\end{equation}
We note that elastic processes involving chemical equilibration are included in the computation of the relaxation time \cite{Sasaki2009}.

Figure \ref{RelaxationTime} shows the relaxation time for the NJL model for $u$ and $s$ quarks as well as the relaxation time for quarks (flavor-blind) and gluons for the DQPM as a function of the temperature $T / T_c$, with $T_c = 200$ MeV for the NJL model and $T_c = 158$ MeV for the DQPM. We can evaluate--in terms of powers of $T$--the behavior of this relaxation time for the NJL model \cite{Zhuang1995}:
\begin{equation}
  T < T_c :
  \begin{cases}
    n \propto e^{-m/T}\\
    \sigma \propto \ \text{const.}
  \end{cases}
  \Rightarrow \tau \propto e^{-T}
\end{equation}
and
\begin{equation}
    T > T_c :
    \begin{cases}
      n \propto T^3\\
      \sigma \propto \ T^{-2}
    \end{cases}
    \Rightarrow \tau \propto T^{-1}.
\end{equation}
The density is proportional to $T^3$ for large temperature where $m \ll T$ in the NJL model, whereas the fast increase of the mass below $T_c$ gives an exponential decrease. The integrated cross sections are roughly constant up to $T_c$ and then are proportional to $1 / T^2$ above $T_c$.

Once one has the relaxation time for the different quark species, it is straight forward to compute all transport coefficients (viscosities and conductivities) in the relaxation time approximation \cite{Sasaki2009}. Obviously all results are strongly $T$ dependent and a systematic study of the $T$ dependence is mandatory.
\vspace{-2mm}
\section{Shear viscosity}
\vskip -3mm
\begin{figure} [t]
  \begin{center}
    \includegraphics[width=8.5cm]{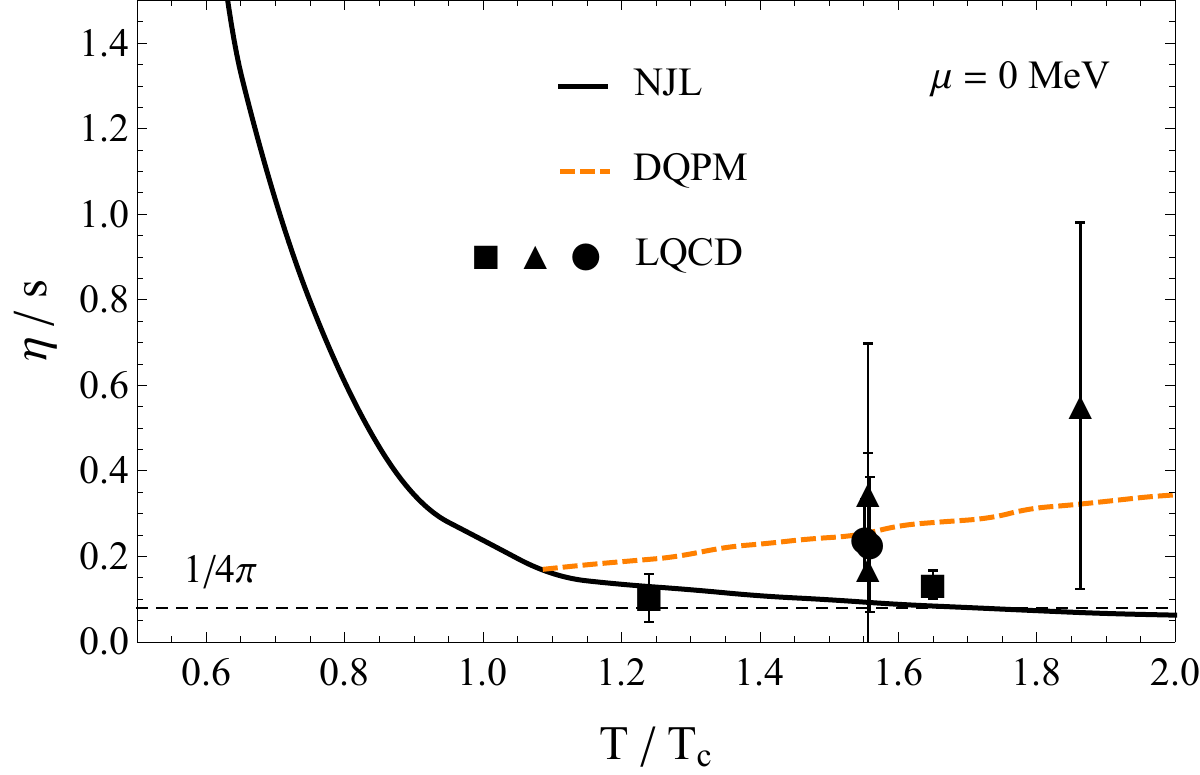}
  \end{center}
  \vskip -5mm
  \caption{(Color online) Shear viscosity to entropy density ratio $\eta / s$ as a function of $T / T_c$ compared to the LQCD data points from Ref. \cite{Meyer2007} (square), \cite{Nakamura2005} (triangle), and \cite{Sakai2007} (circle) and the result from the DQPM.\label{ShearViscosity}}
  \vskip -4mm
\end{figure}

\begin{figure*}
  \begin{center}
    \includegraphics[width=8.5cm]{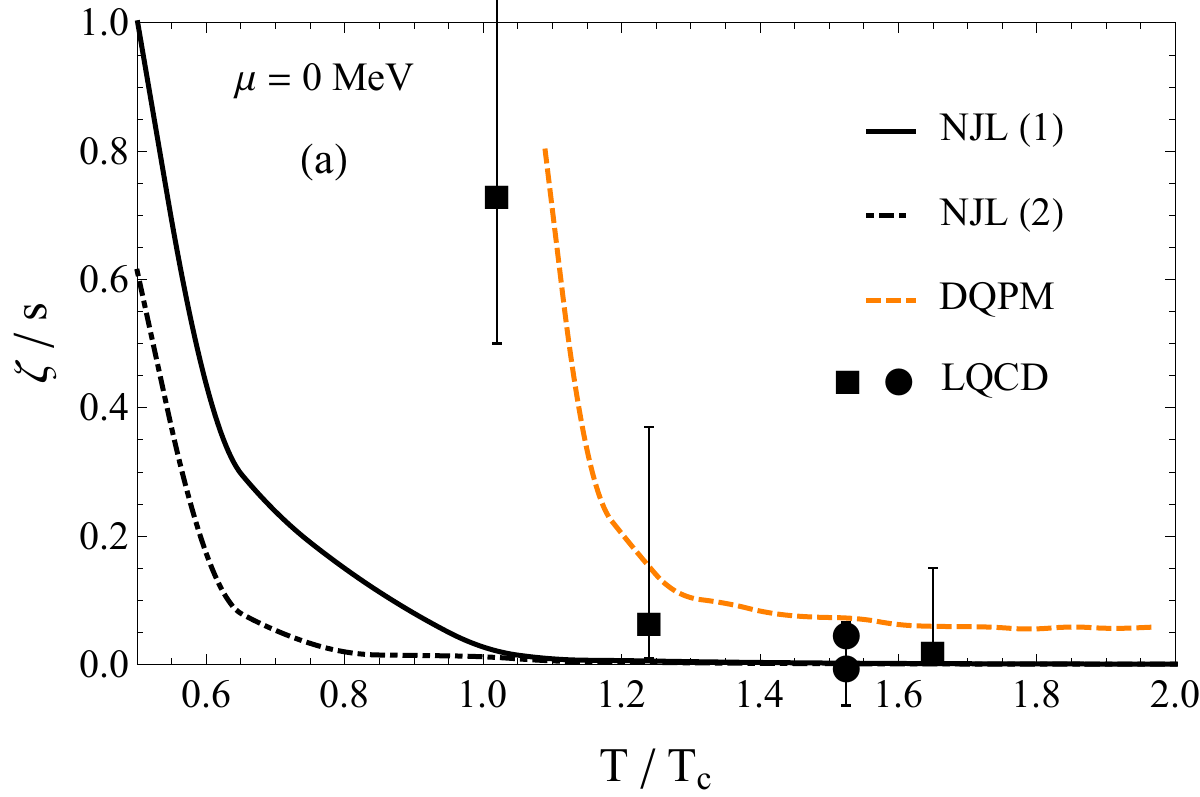} \ \ \
    \includegraphics[width=8.5cm]{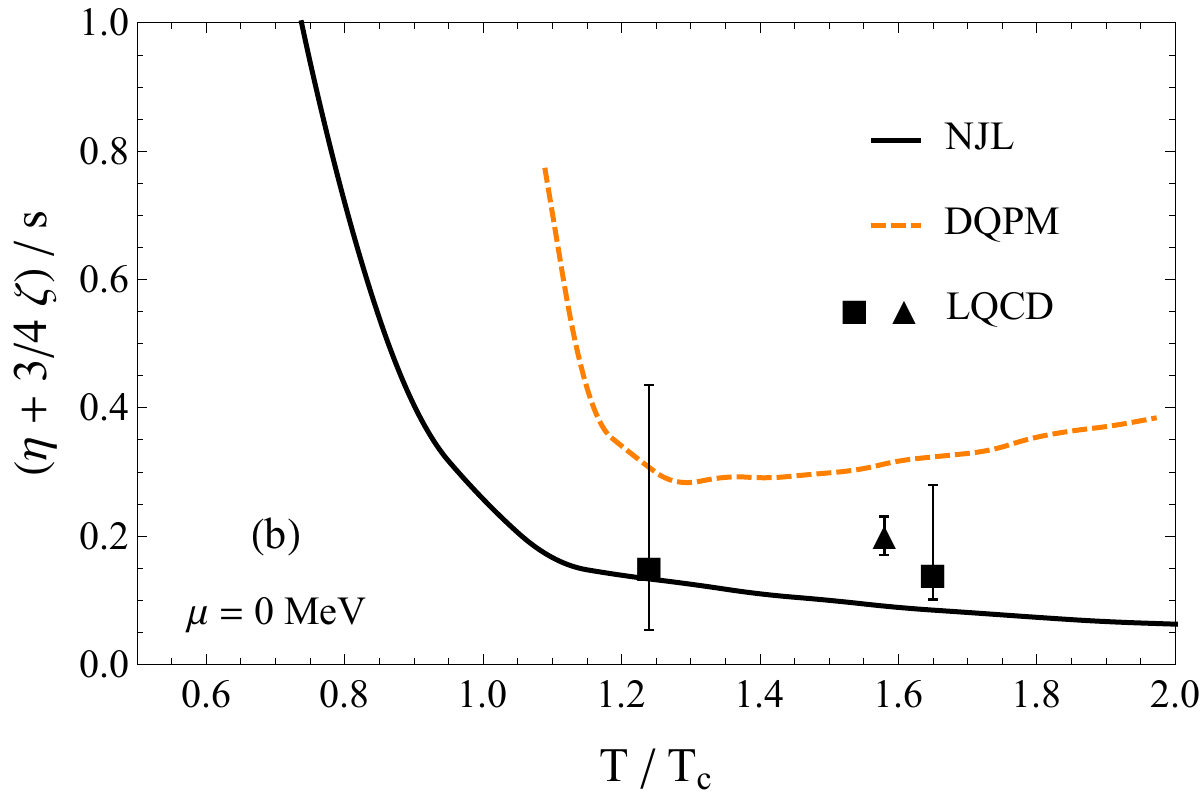}
  \end{center}
  \vskip -5mm
  \caption{(Color online) Bulk viscosity to entropy density ratio $\zeta / s$ (a) compared to the LQCD data points from Ref. \cite{Meyer2008} (square) and \cite{Sakai2007} (circle), and specific sound channel (b) compared to the LQCD data points from Ref. \cite{Meyer2008,Sakai2007} (square) and \cite{Meyer2009} (triangle) as a function of $T/T_c$. For the NJL model, the bulk viscosity to entropy density ratio is computed with the method from Ref. \cite{Chakraborty2010} (1) as well as with the that from Ref. \cite{Sasaki2009} (2). The results from the DQPM are shown by dashed lines.\label{BulkViscosity}}
\end{figure*}

The shear viscosity $\eta$ is related to the transverse motion of a particle during the expansion of a plasma. The calculation of this quantity is very important for the evaluation of physical observables such as the elliptic flow $v_2$. It is defined in the dilute gas approximation for interacting particles as \cite{Chakraborty2010,Bluhm2010}:
\begin{eqnarray}
  \eta(T,\mu)&=&\frac{1}{15 T} g_g           \int_0^\infty \frac{d^3 p}{(2\pi)^3}
  \tau_g f_g \frac{{\bf p}^4}{E_g^2}\\
            & &+\frac{1}{15 T} \frac{g_q}{6} \int_0^\infty \frac{d^3 p}{(2\pi)^3}
  \Bigg[ \sum_q^{u,d,s} \tau_q f_q + \sum_{\bar q}^{\bar u,\bar d,\bar s} \tau_q f_{\bar q} \Bigg] \frac{{\bf p}^4}{E_q^2}. \nonumber
\end{eqnarray}
\vspace{-2mm}

Another approach to compute the shear viscosity is to use the stress-energy tensor and the Green-Kubo formula to extract the viscosity \cite{Ozvenchuk2012b,Plumari2012}. We do not discuss this method in the present study since explicit comparisons of both methods demonstrate that both solutions are rather close (at least in case of approximately isotropic scattering)  \cite{Ozvenchuk2012b,Plumari2012}.

We note that in Ref. \cite{Sasaki2009} the shear and bulk viscosities have been also calculated by restricting to elastic interactions. Their result for the shear viscosity differs from ours only by a Pauli blocking factor $(1-f)$ which is $\approx 1$ for the temperatures of interest. Then the definitions of the shear viscosity from Ref. \cite{Chakraborty2010} and \cite{Sasaki2009} coincide.

We find that the shear viscosity over entropy density ratio $\eta/s(T)$ shows a temperature dependence close to that of the relaxation time $\tau(T)$ in the NJL model. That implies that for $\tau \propto T^{-1}$ for high temperatures one obtains a similar behavior for $\eta/s$ in the NJL model. The order of magnitude of the NJL cross sections finally gives a ratio $\eta/s$ which is in good agreement with lattice QCD data from 1.2 $T_c$ up to 1.5 $T_c$. Nevertheless, the $T^{-1}$ behavior of the viscosity in the NJL implies going beyond the Kovtun-Son-Starinets (KSS) bound \cite{Kovtun2005},
\begin{equation}
  \left(\frac{\eta}{s}\right)_{\text{KSS}} = \frac{1}{4\pi},
\end{equation}which happens around $T \sim 1.7 T_c$. We conclude that the applicability of the NJL model thus should be restricted to temperatures at least below 1.7 $T_c$.

We recall again that the DQPM results agree quite well with LQCD data and give an increasing ratio $\eta/s$ with temperature and approach the pQCD limit at high $T$ as already discussed in Ref. \cite{Ozvenchuk2012b}. One has to note, however, that the lattice data are obtained for pure gauge ($N_f = 0$).

\begin{figure*}
  \begin{center}
    \includegraphics[width=8.5cm]{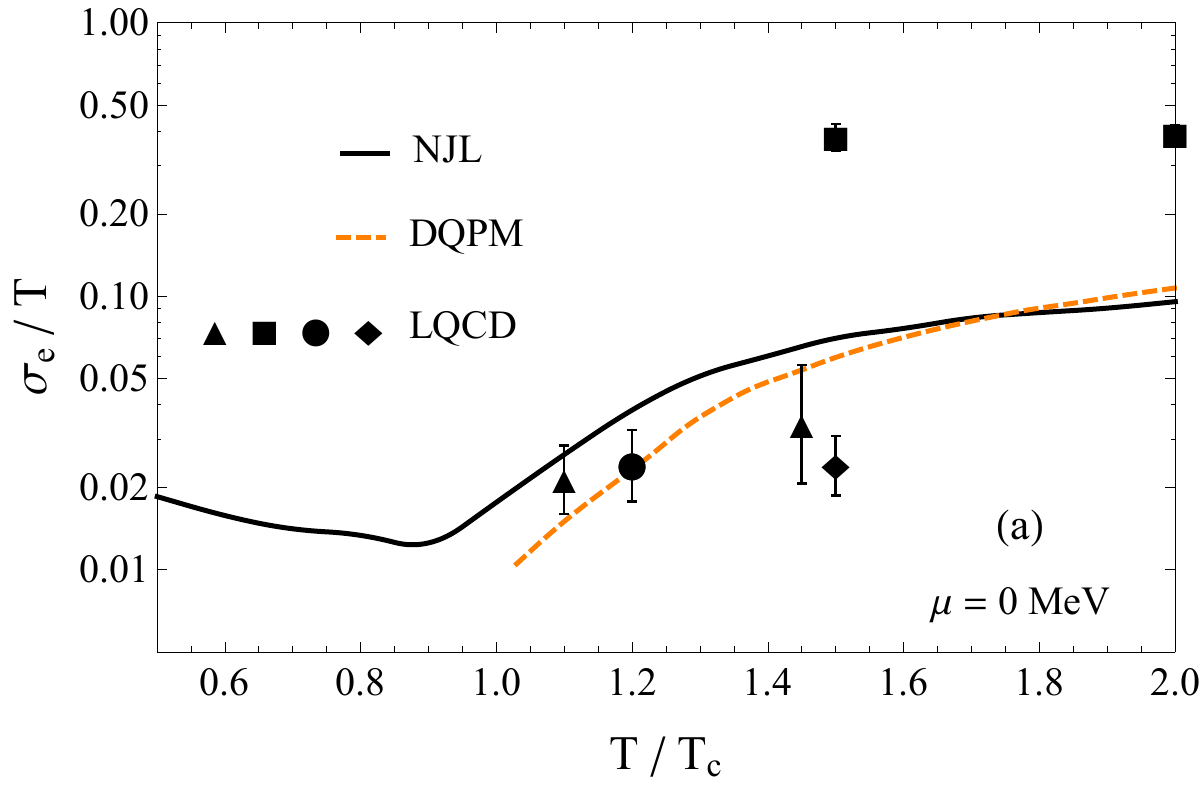} \ \ \
    \includegraphics[width=8.5cm]{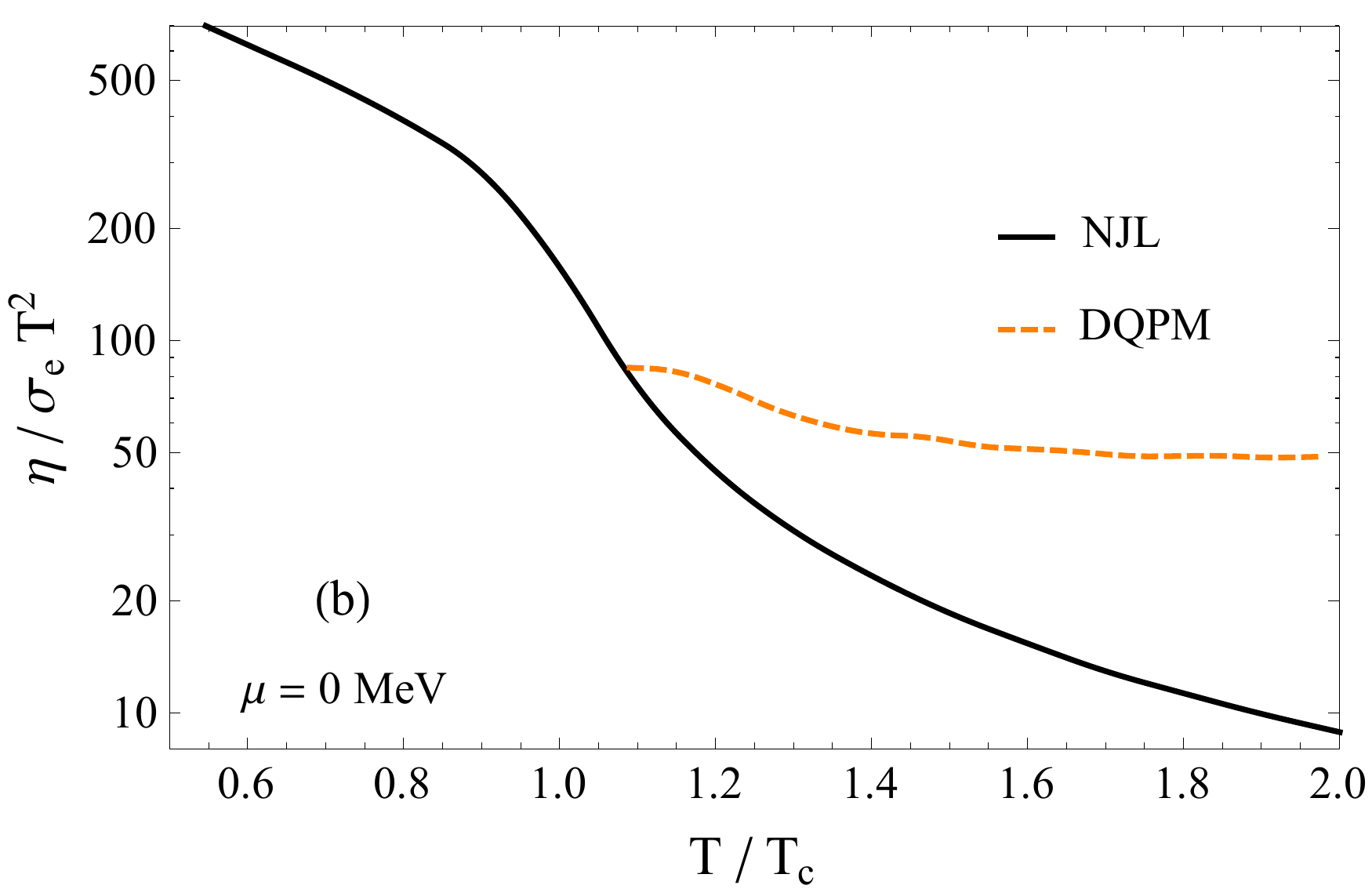}
  \end{center}
  \vskip -5mm
  \caption{(Color online) The electric conductivity over $T$ (a) compared to the LQCD data points from Ref. \cite{Ding2013} (triangle), \cite{Aarts2007} (diamond), \cite{Gupta2003} (square), and \cite{Brandt2013} (circle) and the results from the DQPM (dashed lines); shear viscosity to electrical conductivity ratio $\eta / (\sigma_e T^2)$ (b) as a function of $T/T_c$.\label{ElectricConductivity}}
\end{figure*}
\vspace{-3mm}
\section{Bulk viscosity}
\vskip -3mm
The bulk viscosity defined in Ref. \cite{Chakraborty2010,Bluhm2011} reads in the relaxation time approximation (RTA)
\begin{equation}
  \begin{aligned}
  \zeta(T,\mu) =&\frac{1}{9 T} g_g           \int_0^\infty \frac{d^3 p}{(2\pi)^3} \tau_g f_g\\
  &\times\frac{1}{E_g^2}\Bigg[{\bf p}^2 - 3 c_s^2 \Bigg( E_g^2 - T^2 \frac{d m_g^2}{d T^2} \Bigg) \Bigg]^2\\
              &+\frac{1}{9 T} \frac{g_q}{6} \int_0^\infty \frac{d^3 p}{(2\pi)^3}
  \Bigg[ \sum_q^{u,d,s} \tau_q f_q + \sum_{\bar q}^{\bar u,\bar d,\bar s} \tau_q f_{\bar q} \Bigg]\\
  &\times\frac{1}{E_q^2}\Bigg[{\bf p}^2 - 3 c_s^2 \Bigg( E_q^2 - T^2 \frac{d m_q^2}{d T^2} \Bigg) \Bigg]^2.
  \end{aligned}
\end{equation}

This definition is different from Ref. \cite{Sasaki2009}, even if the same contributions appear (i.e., the mass derivative $\partial m / \partial T$, the energy $E$, the temperature $T$, and the speed of sound squared $c_s^2$). For convenience, we have performed calculations in both approaches in order to get some idea about the differences in results.

The main difference between the RTA approaches of Refs. \cite{Chakraborty2010} and \cite{Sasaki2009} is the definition of the perturbation in the stress-energy tensor $\Delta T^{\mu\nu}$. The RTA approach of Ref. \cite{Chakraborty2010} is based on kinetic theory and allows for both elastic and inelastic scattering (with detailed balance) of an arbitrary number of species; and, most importantly, the viscosities and the equation of state are mutually consistent (the same interactions are used to compute them).


The bulk viscosity [divided by the entropy density $s$] from the NJL model is displayed in Fig. \ref{BulkViscosity}(a) and shows a very different temperature dependence  than $\eta/s$. Indeed, for high temperatures we find the limit $\zeta/s \to 0$. Moreover, the behavior around $T_c$ shows a peak in LQCD as well as in the DQPM. This peak is not seen in the NJL model (or shifted to much lower temperatures). This can be easily explained by the fact that the masses of the degrees of freedom play an important role for the bulk viscosity and the NJL masses do not change sufficiently fast in temperature around $T_c$ to achieve a good agreement with LQCD or the DQPM. When replacing the NJL masses by PNJL masses this does not change the results in this case very much because one has to change the critical temperature as well, and therefore the fast increase of $\zeta/s$ close to $T_c$ cannot be reproduced by any NJL-like model.

\begin{figure*}
  \begin{center}
    \includegraphics[width=8.5cm]{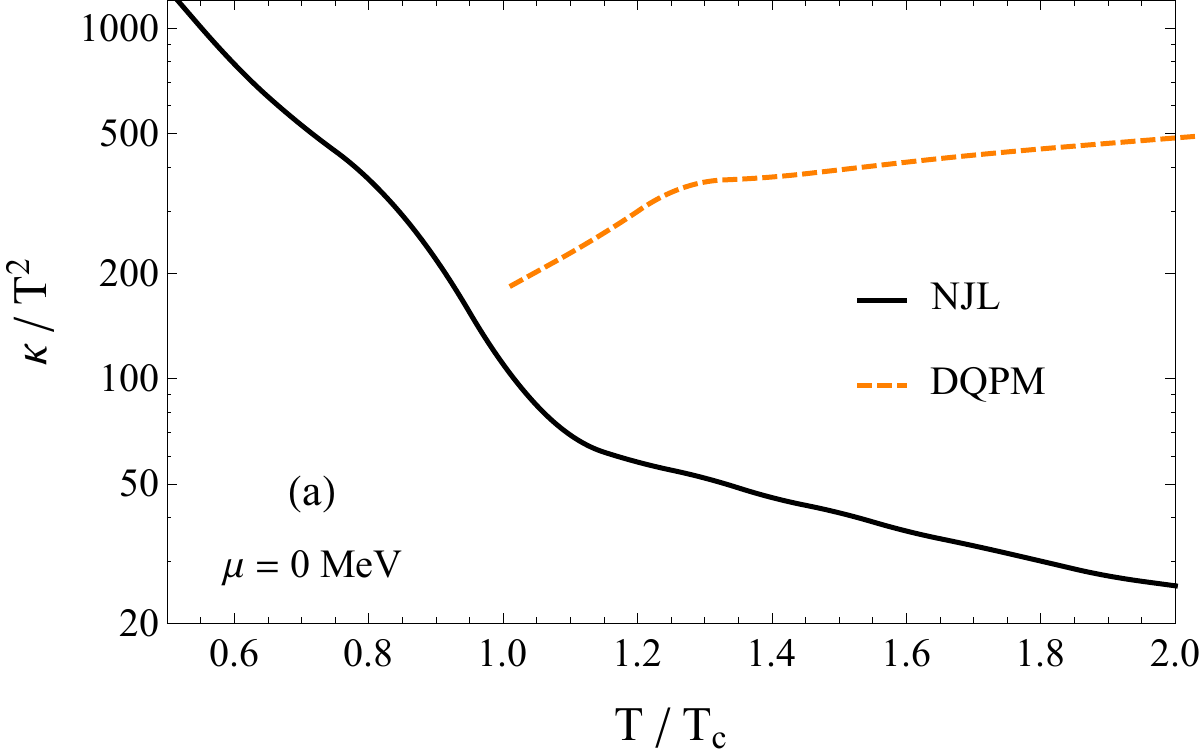} \ \ \
    \includegraphics[width=8.5cm]{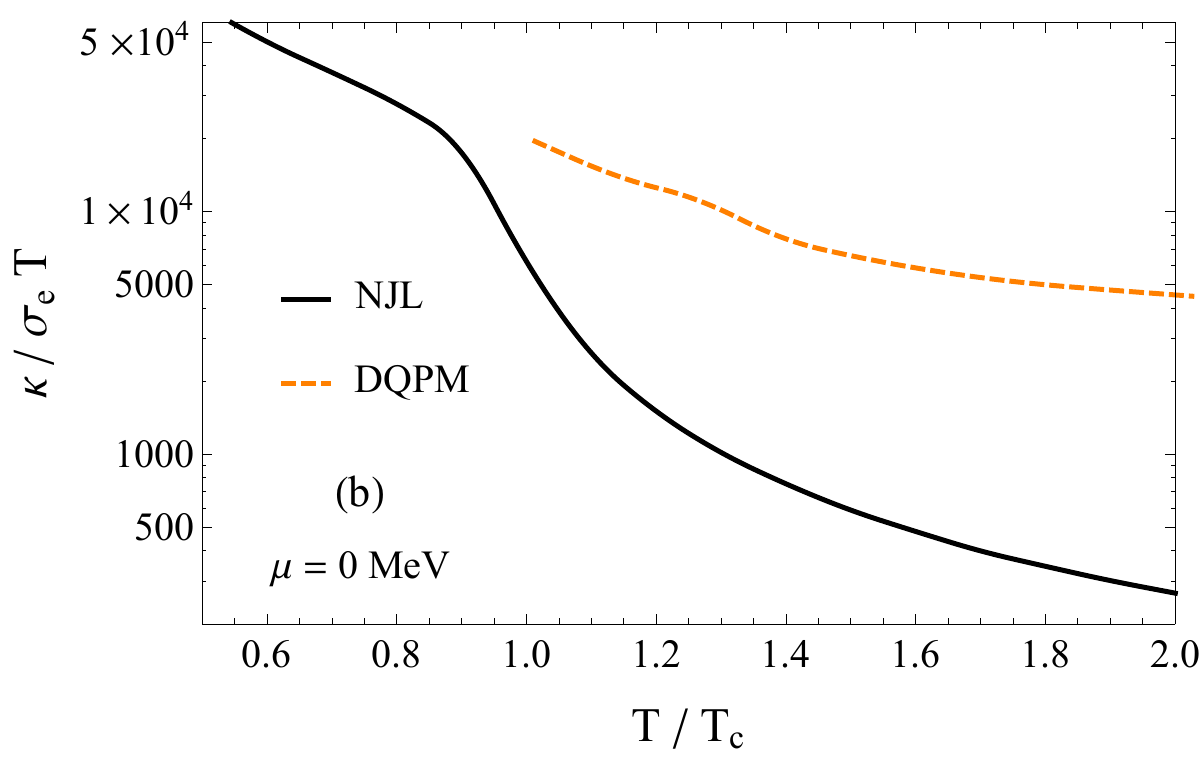}
  \end{center}
  \vskip -5mm
  \caption{(Color online) Heat conductivity over $T^2$ (a) and heat to electrical conductivity $\kappa / (\sigma_e T)$ (b) as a function of $T/T_c$ in the NJL model and the DQPM.\label{HeatConductivity}}
\end{figure*}

In Fig. \ref{BulkViscosity}(b) we also display the specific sound channel $(\eta+3/4\zeta)/s$ which is dominated by $\eta/s$ in the NJL model and provides reasonable results for $T_c < T < 1.7 T_c$ in comparison to LQCD.
\vspace{-4mm}
\section{Electric conductivity}
\vskip -3mm
The electric conductivity for charged particles--known as the Drude-Lorentz conductivity for a classical gas--is defined as \cite{Reif1965,Cassing2013}
\begin{equation}
  \sigma_e(T,\mu) = \sum_q \frac{e_q^2 \ n_q(T,\mu) \ \tau_q(T,\mu)}{m_q(T,\mu)},
\end{equation}
with $q = u,d,s,\bar u,\bar d,\bar s$, and the electric charge of quarks,
\begin{equation}
  e_q^2 = \frac{4\pi}{137} q^2,
\end{equation}
with $q = +2/3$ or $-1/3$ denoting the quark electric charge fraction.

Figure \ref{ElectricConductivity}(a) shows that for the DQPM as well as the NJL model the dimensionless ratio of the electric conductivity over $T$ behaves approximately linearly in $T$ for $T \geq T_c$ up to about 2 $T_c$ \cite{Cassing2013}. Both results are in a reasonable agreement with the present lattice QCD results although there is quite some uncertainty in the LQCD extrapolations.

In Fig. \ref{ElectricConductivity}(b) we also show the dimensionless ratio $\eta / (\sigma_e T^2)$ which is independent on the relaxation time $\tau$. This ratio is roughly constant in the DQPM for $1.4 T_c < T < 2 T_c$; however, it drops rapidly in the NJL model with temperature $T$.

\section{Heat conductivity}
\vskip -3mm
The heat conductivity $\kappa$ is another quantity of interest that describes the heat flow in interacting systems \cite{Israel1979,deGroot1980} and only recently has regained interest in the context of relativistic heavy-ion collisions \cite{Denicol2012,Greif2013}.

The heat conductivity for charged particles is defined using the specific heat $c_V$ and the relaxation time \cite{Heiselberg1993}:
\begin{equation}
  \kappa(T,\mu) = \frac{1}{3} v_{\text{rel}} \ c_V(T,\mu) \ \sum_f \tau_f(T,\mu),
  \label{heatcond}
\end{equation}
with $f = u,d,s,\bar u,\bar d,\bar s$. For our purpose we assume that $v_{\text{rel}} \simeq 1$ in the NJL model because the masses of quarks decrease whereas the mean momentum increases with temperature $T$. Since all quantities entering in Eq. \eqref{heatcond} have been specified before for the NJL model and the DQPM we directly proceed with the actual results.

Figure \ref{HeatConductivity}(a) displays the dimensionless quantity $\kappa / T^2$ for both models. Whereas the DQPM shows a slightly rising ratio for $T_c < T < 2 T_c$ the NJL model predicts a rapid decrease with $T$ for $T > T_c$.

In Fig. \ref{HeatConductivity}(b) we show the results of both models for the dimensionless ratio $\kappa / (\sigma_e T)$, which no longer depends on the relaxation time $\tau$. For $T > T_c$ the results differ by more than an order of magnitude. Unfortunately, there are no lattice QCD results that might specify the quality of the predictions from both models.

\section{Conclusion}
\vskip -3mm
In this study we have calculated thermodynamic properties of the NJL model for three flavors ($u,d,s$) such as the energy density, entropy density, pressure, sound velocity, specific heat, etc. as a function of temperature $T$ up to a few times the critical temperature $T_c$. Furthermore, we have calculated the shear $\eta$ and bulk $\zeta$ viscosity as well as the electric $\sigma_e$ and heat conductivity $\kappa$ as a function of $T$ and compared to corresponding results from the DQPM, from the PNJL model, and from lattice QCD results when available. Note, however, that most of them have been calculated in pure gauge ($N_f=0$).

We recall that the NJL Lagrangian parameters (as well as a momentum cutoff) are essentially fixed at $T=0$ by the pion (kaon) decay constant and pion (kaon) mass whereas the PNJL model is improved especially around the critical temperature $T_c$ by adding a Polyakov loop potential (fitted to reproduce specific lattice data of the HOT-QCD Collaboration). Both models have in common that the Lagrangians share the symmetries of QCD at low temperature. On the other hand the DQPM is based on a two-particle irreducible approach for complex ``resummed'' propagators and ``resummed'' couplings especially tuned to lattice QCD results from the Wuppertal-Budapest Collaboration for the entropy density $s(T)$ above temperatures of 140 MeV. In this respect the DQPM does not have a well defined effective Lagrangian in the low-temperature limit that might allow for an explicit calculation of cross sections; however, the interaction rates of the partons are included in the width of the spectral functions such that the various transport coefficients can be calculated in a straight forward manner. Unfortunately, the lattice QCD results from the two collaborations differ substantially due to the diffenrent implementations of the action and bare quark masses. Accordingly, the various comparisons have to be taken with care and not on an absolute scale.

We have seen that the NJL model does not reproduce the lattice QCD equations of state; first of all because the number of degrees of freedom is not sufficient in the vicinity of $T_c$, but also because there is no explicit control on gluon degrees of freedom and their contribution to the thermodynamic potential. However, the results for the thermodynamic quantities $s(T), P(T)$, etc. approximately decouple from the transport properties, which appear reasonable in the vicinity of $T_c$ up to about 1.5 $T_c$, whereas the NJL fails to describe the transport coefficients in the high-temperature phase for $T \geq 1.5 T_c$. Indeed, as confirmed by the relaxation time $\tau$ or by the shear viscosity to entropy density ratio $\eta / s$, the NJL model overestimates the interactions between the low-mass constituent quarks in the medium at high temperature and lacks the gluon dynamics. The PNJL model performs better for the entropy and energy density due the introduction of an explicit Polyakov-loop potential that describes the pressure from the gluonic degrees of freedom as a function of temperature and leads to a narrower cross over region around $T_c$.

It is interesting to see that, despite the small number of parameters of the NJL model, one can easily describe the thermodynamical properties as well as the transport coefficients within some deviation from LQCD results in equilibrium. Other models like the DQPM are designed to match LQCD results and are very useful to describe a realistic plasma with the right number of degrees of freedom, but this is not an effective Lagrangian model where one can directly extract cross sections. We note in passing that this task can be solved to some extent within the PHSD transport approach which has been shown to mach the DQPM results in thermodynamic equilibrium \cite{Ozvenchuk2012a}. In any cases neither the (P)NJL model nor the DQPM can describe explicitly the dynamics of confinement.

The important message we want to point out here is that --although there is no perfect model to describe the physics of the QGP-- there are effective approaches that reproduce thermodynamic as well as transport properties of QCD in equilibrium, as shown by the explicit comparison to LQCD results. Furthermore, we have pointed out quantitatively the limitations of the models with respect to their applicability in energy density (or temperature) and provided ratios for transport coefficients that in principle can be well controlled by LQCD calculations. Further studies on the NJL model in a box, for equilibration times, strangeness dynamics, and a mixed phase (in analogy to Ref. \cite{Ozvenchuk2012a,Ozvenchuk2012b}), are in progress.

\section{Acknowledgment}
\vskip -3mm
We thank H. Hansen, D. Rischke, C. Sasaki, and N. N. Scoccola for illuminating discussions. We are also grateful to O. Linnyk and V. Ozvenchuk for providing various LQCD data points. We also acknowledge support through the ``HIC for FAIR" framework of the ``LOEWE" program. The computational resources have been provided by the LOEWE-CSC.

  \bibliography{biblio}

\end{document}